\documentclass[a4paper,twocolumn,11pt,accepted=2023-10-31]{quantumarticle}
\pdfoutput=1
\usepackage{amsmath}
\usepackage{amssymb}
\usepackage{braket}
\usepackage{hyphenat}
\usepackage{amsthm}
\usepackage{color}
\usepackage{bbm}
\usepackage{bm}
\usepackage{upgreek}
\usepackage[numbers]{natbib}
\usepackage{graphicx}
\usepackage{pdftexcmds}
\usepackage{mathrsfs}
\usepackage{enumitem}
\usepackage{array}
\usepackage[hyperfootnotes=true]{hyperref}
\usepackage[figure]{hypcap}
\usepackage{tikz} 
\usepackage{mathtools}
\usepackage{algorithm}
\floatname{algorithm}{Box}
\usepackage{algpseudocode}
\usepackage{soul}
\usepackage{adjustbox}
\usepackage{array}
\usepackage{comment}

\DeclareFontFamily{U}{cbgreek}{}
\DeclareFontShape{U}{cbgreek}{m}{n}{
        <-6>    grmn0500
        <6-7>   grmn0600
        <7-8>   grmn0700
        <8-9>   grmn0800
        <9-10>  grmn0900
        <10-12> grmn1000
        <12-17> grmn1200
        <17->   grmn1728
      }{}
\DeclareFontShape{U}{cbgreek}{bx}{n}{
        <-6>    grxn0500
        <6-7>   grxn0600
        <7-8>   grxn0700
        <8-9>   grxn0800
        <9-10>  grxn0900
        <10-12> grxn1000
        <12-17> grxn1200
        <17->   grxn1728
      }{}

\makeatletter
\newcommand{\normalorbold}{%
  \ifnum\pdf@strcmp{\math@version}{bold}=\z@ bx\else m\fi
}
\makeatother

\def\gbm#1{{\let\pi\uppi \let\phi\upphi \let\lambda\uplambda \let\mu\upmu \let\rho\uprho \let\sigma\upsigma \let\tau\uptau \let\theta\uptheta \let\eta\upeta \bm{#1}}}

\newcommand*{\bbC}{\mathbb{C}}
\newcommand*{\bbR}{\mathbb{R}}

\newcommand*{\bbZ}{\mathbb{Z}}

\newcommand*{\cE}{\mathcal{E}}

\newcommand*{\cM}{\mathcal{M}}

\newcommand*{\cX}{\mathcal{X}}

\newcolumntype{L}{>{\centering\arraybackslash}m{2cm}}

\newcommand*{\kbra}[2]{\ket{#1\vphantom{#1 #2}}\bra{#2\vphantom{#1 #2}}}

\newcommand*{\proj}[1]{\kbra{#1}{#1}}

\newcommand*{\tr}{\mathrm{Tr}}
\newcommand*{\id}{\mathrm{id}}
\newcommand*{\fr}[2]{\frac{#1}{#2}}

\newcommand{\vect}[1]{\mathbf{#1}}

\newcommand{\be}{\begin{equation}}
\newcommand{\ee}{\end{equation}}
\newcommand{\n}{\textendash}
\newcommand{\m}{\textemdash}

\newcommand{\kB}{k_\text{B}}
\newcommand{\St}{S}
\newcommand{\SSet}{\mathcal{S}} 
\newcommand{\Obs}{O}
\newcommand{\littleObs}{o}

\newcommand{\wlambda}[1][\lambda]{w^{(#1)}}
\newcommand{\EP}{\Sigma}

\newcommand{\entropyrate}{s_\text{vN}}
\newcommand{\myopicentropyrate}{s}
\newcommand{\wideal}{w^\text{ideal}}
\newcommand{\EE}{\mathcal{E}}

\hypersetup{
    pdftoolbar=true,        % show Acrobat’s toolbar?
    pdfmenubar=true,        % show Acrobat’s menu?
    pdffitwindow=false,     % window fit to page when opened
    pdfstartview={FitH},    % fits the width of the page to the window
}

\begin{document}
\title{Engines for predictive work extraction from memoryful quantum stochastic processes}
\date{December 11, 2022}
%\date{\today}

\author{Ruo Cheng Huang}
\email{ruocheng001@e.ntu.edu.sg}
\affiliation{Nanyang Quantum Hub, School of Physical and Mathematical Sciences, Nanyang Technological University, Singapore
}

\author{Paul M.\ Riechers}
\email{pmriechers@gmail.com}
\orcid{0000-0002-0135-3778}
\affiliation{Nanyang Quantum Hub, School of Physical and Mathematical Sciences, Nanyang Technological University, Singapore
}
\affiliation{Beyond Institute for Theoretical Science (BITS), San Francisco, CA, USA}

\author{Mile Gu}
\email{mgu@quantumcomplexity.org}
\affiliation{Nanyang Quantum Hub, School of Physical and Mathematical Sciences, Nanyang Technological University, Singapore
}
\affiliation{Centre for Quantum Technologies, National University of Singapore, 3 Science Drive 2, Singapore}
\affiliation{MajuLab, CNRS-UNS-NUS-NTU International Joint Research Unit, UMI 3654, 117543, Singapore}
\author{Varun Narasimhachar}
\email{varun.achar@gmail.com}
\affiliation{Nanyang Quantum Hub, School of Physical and Mathematical Sciences, Nanyang Technological University, Singapore
}
\affiliation{A*STAR Quantum Innovation Centre (Q.InC), Institute of High Performance Computing (IHPC), Agency for Science, Technology and Research (A*STAR), 1 Fusionopolis Way, Republic of Singapore 138632}

\begin{abstract}
    Quantum information\hyp processing techniques enable work extraction from a system's inherently quantum features, in addition to the classical free energy it contains. Meanwhile, the science of computational mechanics affords tools for the predictive modeling of non\hyp Markovian classical and quantum stochastic processes. We combine tools from these two sciences to develop a technique for predictive work extraction from non\hyp Markovian stochastic processes with quantum outputs. We demonstrate that this technique can extract more work than non\hyp predictive quantum work extraction protocols, on the one hand, and predictive work extraction without quantum information processing, on the other. We discover a phase transition in the efficacy of memory for work extraction from quantum processes, which is without classical precedent. Our work opens up the prospect of machines that harness environmental free energy in an essentially quantum, essentially time\hyp varying form.
\end{abstract}

\maketitle

%\section{Introduction}
% \noindent\textbf{The quantum thermodynamic revolution}\m The emerging field of quantum thermodynamics, which lies at the interface of quantum information and thermodynamics, has seen advances on both foundational and applied fronts [About work extraction from quantum states \cite{SSP14} and other literature] One productive approach has been to frame quantum thermodynamics as a \emph{resource theory}: a framework where quantum states away from equilibrium are characterized as resources powering operational tasks such as work extraction and cooling. But the rich interplay of quantum mechanics and thermodynamics cannot entirely be captured by the \emph{static} resources in the form of non\hyp equilibrium states\m the complexity of quantum thermodynamic processes is \emph{dynamical} in nature. Recent advances have initiated the treatment of channels and more complex quantum processes as resources in their own right, but the results thus far are abstract and removed from concrete physical settings. In particular, existing models of engines that can extract inherently quantum\hyp mechanical forms of free energy  are not equipped to handle the complex dynamical forms of fuel that we consider in the present work.\vspace{2ex}

\section{Introduction}
In the earliest heat engines, a combustible fuel was burned to maintain a temperature gradient between hot and cold heat reservoirs. The second law of thermodynamics holds that no engine can sustainably function with a single reservoir \cite{schroeder1999introduction,blundell2009concepts,sethna2021statistical,callen1991thermodynamics}. While thought experiments such as Maxwell's demon and Szilard's engine initially appear to defy this law~\cite{szilard1964decrease}, a more complete understanding of thermodynamics resolved the apparent paradox: the resource powering the engine need not be a temperature gradient, but may be \emph{any} form of free energy\m even information \cite{landauer1961irreversibility,bennett1982thermodynamics,mandal2012work,mandal2013maxwell, parrondo2015thermodynamics}. The emerging field of quantum thermodynamics has continued to expand the scope of ``fuel'' to increasingly general forms of free energy \cite{goold2016role,binder2018thermodynamics}. There has been both theoretical and experimental advancement in constructing engines that can harness the free energy locked up in quantum coherence, over and above classical free energy \cite{SSP14,aaberg2014catalytic,korzekwa2016extraction,garner2017thermodynamics,lostaglio2018thermodynamic}.

The story does not stop there\m in addition to \emph{static} fuel, there is also a \emph{dynamical} fuel\hyp like resource embodied by complex thermodynamic processes. The framework of \emph{computational mechanics} in complexity science offers powerful techniques for the characterization and manipulation of stochastic processes. The future behaviour of such a process in general cannot be known perfectly using data from its past. Nevertheless, temporal correlations, i.e.,\ patterns in a process's behaviour over time, enable prediction. These correlations may even be \emph{non\hyp Markovian}, whereby the future of a process depend not only on its present, but also on its distant past. Epsilon machines and their quantum extensions \cite{crutchfield1989inferring,shalizi2001computational,crutchfield2012between} perform memory\hyp optimal predictive modelling of stochastic processes. %, while p
Pattern extractors \cite{garner2017thermodynamics, Boyd17PRE} 
%are thermodynamic machines that use 
leverage
prediction to extract useful work from the classical free energy present in temporal patterns, exchanging heat with a single bath. However, these predictive engines are not equipped to harness the free energy locked up in quantum degrees of freedom.
%\vspace{2ex}

%\noindent\emph{Our contributions}\m 
%In this letter, 

\subsection{Our contributions}

Here, 
we develop the theoretical prototype for a \emph{predictive quantum engine}: a machine that charges a battery by feeding on a multipartite quantum system whose parts are temporally correlated via a classical stochastic process. In other words, the engine's fuel is a \emph{classical stochastic process with quantum outputs}. It can extract free energy beyond what is accessible to current quantum engines or classical predictive engines. We present a systematic construction of such an engine for arbitrary classical processes and quantum output states.
We illustrate its application 
on example stochastic processes
%via the \emph{perturbed coin} process\m a stochastic sequence 
of %classically 
correlated non-orthogonal qubits. 
%\VN{``non-orthogonal qubits'' is a somewhat unsatisfactory / inaccurate description to me. Flagging this in case we want to rephrase it.}%
%\PMR{I'm fine with it, but open to suggestions.}
(Fig.~\ref{fig:perturb_coin}).  
We also use this test case to benchmark the performance of our engine against various alternatives, including one without coherent quantum information processing, and one without predictive functionality.
Our predictive quantum engine 
%is shown to 
outperforms these alternatives in terms of work output. 
%Furthermore, we evaluate the engine's long-term performance on fuel processes with different degrees of temporal correlations, and find the work yield to increase with such correlations. 
We show 
that parametrized processes
of 
correlated non-orthogonal
quantum outputs exhibit phase boundaries between parametric regions where memory of past observations can and cannot enhance the work yield---despite the apparently smooth change of memoryful correlations in the process across this boundary. 
The sudden lack of
memory advantage is thus fundamentally thermodynamic (since prediction per se 
has more freedom than during
work extraction)
%would not suffer
%but work extraction does) 
and fundamentally quantum
(since classical engines can exploit all the process’s inherent memory).
Finally, we generalize the Information Processing Second Law (IPSL) to the quantum regime and derive the fundamental bounds on a quantum pattern engine's performance. 
We begin by introducing stochastic processes as a source of quantum states that will fuel our engine in Section \ref{sec:fuel}.
Section \ref{sec:synchronize} 
shows how general interactions with the correlated quantum states induce belief states that enable optimal prediction and work extraction.
We then illustrate how a theoretical pattern engine can be constructed in Section \ref{sec:construct}. The performance of the engine on an example process is then evaluated in Section \ref{sec:example}.
This leads to the discovery and explanation of the aforementioned phase transition in the efficacy of memory. 
The fundamental limit of a pattern engine is then derived in Section \ref{sec:FunLimits}, and finally we draw our conclusion in Section \ref{sec:discussion}.
\subsection{What fuels our engine}
\label{sec:fuel}
\begin{figure}[t] % diagram for the perturbed coin process
\includegraphics[width=\columnwidth]{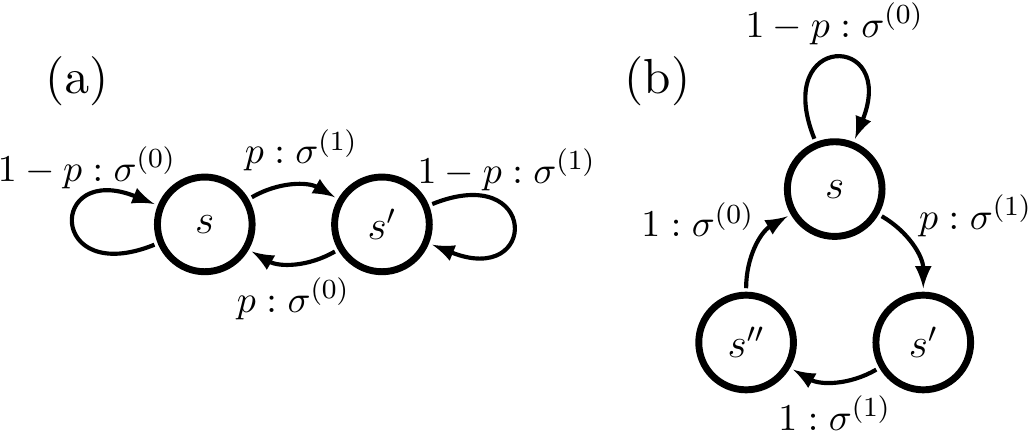}
\caption{Latent-state sources of correlated quantum processes. %Stochastic graph, e
Each arrow represents a transition between latent states; the label $p:\sigma^{(x)}$ indicates that the transition happens with probability $p$ and
produces a quantum state $\sigma^{(x)}$.
%emits a symbol $x$. 
%For classical case, $\sigma^{(0)}$ and $\sigma^{(1)}$ are perfectly distinguishable states.
(a) Perturbed-coin process. (b) 2-1 golden-mean process.}
\label{fig:perturb_coin}
\end{figure}

Typically in quantum information theory,
sources of quantum states are assumed 
to be memoryless, 
producing independent and identically distributed (IID) quantum states at each discrete time~\cite{Nielsen02Quantum, Stras17Quantum}.
%However, nature is animated %replete 
On the contrary, nature abounds
with rich dynamics.
To go beyond the IID paradigm, we
consider general \emph{finite-state sources
of quantum states},
which can create highly nontrivial correlations across time.  
Some simple examples are depicted in Fig.~\ref{fig:perturb_coin}
%Memoryfull structure 
Temporally patterned
quantum outputs provide fuel 
for any entity capable of predicting them.
%Supplementary Material 
Appendix \ref{app:FreeEnergyDecomp}
shows that any change in \emph{total correlation} among outputs,
as quantified by quantum relative entropy,
%contributes additively to
%is an additive contribution to the harvestable 
changes 
nonequilibrium free energy
proportionally.
Correlations are therefore a source of free energy.

These memoryful sources of quantum states 
can be represented by a 
 hidden Markov model (HMM) $\mathcal{M} = \Bigl( \SSet, \bigl( \sigma^{(x)} \bigr)_{x \in \mathcal{X}}, \bigl(T^{(x)}\bigr)_{x \in \mathcal{X}} \Bigr)$.
Here, $\SSet$ is the HMM's set of classical latent states.
The random variable representing the latent state at time $t$ shall be denoted by $\St_t$. 
%\VN{We should mention that latent states are classical}
The transition-matrix
element
$T^{(x)}_{s,s'} 
= \Pr(\St_t = s', X_t = x | \St_{t-1} = s)$ 
represents the probability of transitioning from latent state $s$ to $s'$ and emitting the $d$-dimensional quantum state $\sigma^{(x)}$.
%\VN{This description of $\sigma$ is clever for its parsimony, but probably too late. I think we should mention what it is together with the rest of the HMM. Space is not a concern for Quant}
%\PMR{We're still introducing the HMM here. I think it just takes time.}

For simple HMMs, the memoryful transition structure can be %usefully 
visualized as an annotated directed graph, as in Fig.~\ref{fig:perturb_coin}.  
In this graphical representation, nodes correspond to latent states while the directed edges correspond to latent-state-to-state transitions that produce a certain quantum output with a prescribed probability. 

In the example of Fig.~\ref{fig:perturb_coin}(a), the `biased-coin process' has only two latent states $s$ and $s'$.  At each timestep, the latent state switches with probability $p$.  The quantum output is $\sigma^{(0)}$ whenever the resultant state is $s$; 
The quantum output is $\sigma^{(1)}$ whenever the resultant state is $s'$.
As the switching-probability $p$ approaches 1,
the process approaches a period-two output, alternating between the two quantum states $\dots \sigma^{(0)} \otimes \sigma^{(1)} \otimes \sigma^{(0)} \otimes \sigma^{(1)} \dots$.
At the other extreme, as $p$ approaches 0,
the latent state remains the same for increasingly long epochs 
that produce long strings of the same quantum output
$\dots \sigma^{(0)} \otimes \sigma^{(0)} \otimes \sigma^{(0)} \otimes \sigma^{(0)} \dots$
or
$\dots \sigma^{(1)} \otimes \sigma^{(1)} \otimes \sigma^{(1)} \otimes \sigma^{(1)} \dots$
switching between the two behaviors only rarely.
More generally, for any $p \in (0,1)$,
the process interpolates between these two extreme behaviors.

More complicated 
HMMs can generate more complex
memoryful structure 
in the quantum-output process.
See Fig.~\ref{fig:perturb_coin}(b) 
for a hint of this possible richness.

%In the example of Fig.~\ref{fig:perturb_coin}(a), the process at any time can be in either state $s$ or $s'$, the probability of the next output will then depend on the state as shown by the arrows. In this particular example, when $p=1$, then at every time step, the process will deterministically transition from $s$ to $s'$($s'$ to $s$) emitting quantum state $\sigma^{(1)}$($\sigma^{(0)}$). When $p=0$ on the other hand, the process becomes a constant process that only outputs quantum state $\sigma^{(0)}$ or $\sigma^{(1)}$ depending on whether the system is in $s$ or $s'$. 

The HMM specifies the statistics of the non-Markovian classical variables $X_t$ across time, which, in turn, induce the 
quantum outputs indexed by $x \in \mathcal{X}$.
The quantum output process is described by the (formal) density operator
\be\label{eqfuel}
\rho_{\overleftrightarrow A}=\sum_{\overleftrightarrow x} \Pr \left(\overleftrightarrow x\right)\bigotimes_{t\in\bbZ}\sigma^{(x_t)}_{A_t},
\ee
where each time step $t$ is associated with a unique elementary physical system $A_t$, and $\overleftrightarrow x = \dots x_{-1}x_0x_1\dots$ denotes a bi\hyp infinite string over $\cX$.
The joint quantum state is separable amongst the $A_t$s, but 
can have non-classical correlations in the form of discord~\cite{Modi10Unified}.
These memoryful quantum sources 
generalize the kindred
`classically controlled qubit sources' of Ref.~\cite{venegas2020measurement}.

\begin{figure} % schematic diagram of an engine
\begin{center}
\includegraphics[width=\columnwidth]{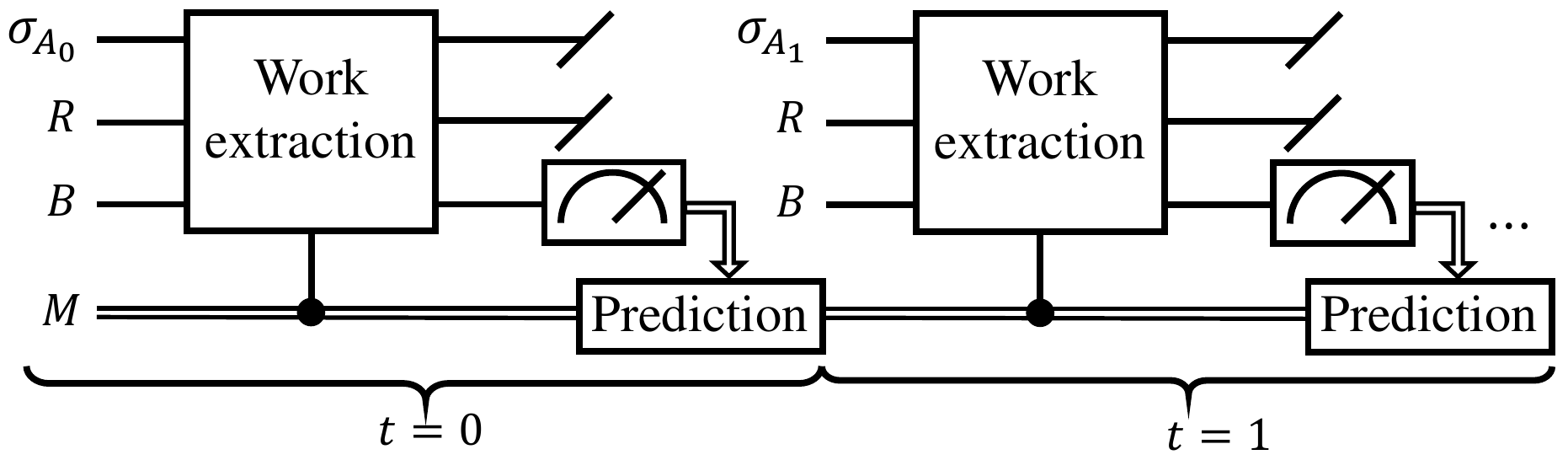}
\caption{Schematic diagram of a quantum-pattern engine.  At each time step, the process will take a quantum system, $\sigma_{A_t}$, from the ``fuel" tape, reservoir qudit, $R$, battery, $B$, and memory, $M$, as input. 
The `Work extraction' box
should be interpreted as 
a memory-dependent unitary. 
States of battery and memory are recycled.
}
\label{fig:schem_eng}
\end{center}
\end{figure}

The time\hyp indexed sequence of stochastically\hyp generated quantum outputs $\sigma^{(x_t)}_{A_t}$ of such a process acts as a ``fuel tape'' that is fed to the engine (Fig.~\ref{fig:schem_eng}) one $A_t$ at a time, in a temporal order. 
If the quantum states could all be stored for a later time,
then it would be straightforward to extract all nonequilibrium addition to free energy from the correlated quantum tape, by acting on the full sequence all at once~\cite{SSP14}, since the multipartite system could then be treated as a single quantum system.
Rather, we consider a setting where each elementary physical system
has an immediate expiration date\m the engine can 
interact with 
%the present elementary system 
$A_t$
at the present time $t$,
and then never again.
This time-locality restriction is 
%reminiscent of 
shared with
the information-ratchet~\cite{mandal2012work, Boyd17PRE}
and repeated-interaction~\cite{Stras17Quantum} frameworks.
The challenge here is to extract maximal free energy
when forced to act only locally and sequentially on the correlated quantum pattern.

Except for being constrained by a stationary finite\hyp state source, the fuel process can be arbitrary in its alphabet size, statistics, and quantum outputs' finite dimensionality and form\m $\sigma^{(x)}$ could be a pure or mixed state over any number of quantum degrees of freedom. Notice that here the form of the set of $\{\sigma^{(x)}\}_{x\in\mathcal{X}}$ does not vary with time, hence the lack of $t$ index. 
%For brevity of discussion, we also omit the system label $A_t$ hereon unless required.
To simplify notation,
we will omit the system label $A_t$ unless required.

We assume that the %form \eqref{eqfuel} 
source
of the fuel tape is known exactly. This entails a complete knowledge of the underlying classical statistics and of the indexed set of quantum states $\{\sigma^{(x)}\}_{x\in\mathcal{X}}$, but not of which specific string $\overleftrightarrow x$ is instantiated. %\VN{Would this paragraph fit better before
%the previous paragraph?}
%\PMR{Its current placement motivates the next section on synchronization. }

%We are now ready for our first result: the general construction of a quantum pattern engine for any given fuel process.\vspace{2ex}

%The quantum systems $A_t$ are classically correlated; they can be generated by a memoryfull (possibly non\hyp Markovian) source that can be represented by a finite\hyp state hidden Markov model (HMM)
%$\mathcal{M} = \bigl( \SSet, \{ \sigma^{(x)} \}_{x \in \mathcal{X}}, \{T^{(x)}\}_{x \in \mathcal{X}} \bigr)$. %[Check the use of curly vs.\ round brackets: should it be consistent between T and sigma?].
%Here, $\SSet$ is the HMM's set of latent states.
%If the generator is in latent state $s$ then, with probability $T_{s,s'}^{(x)} = \Pr(\St_{t+1} = s', X_t = x | \St_t = s)$, it will transition to state $s'$ while 
%preparing and transmitting the physical state $\sigma^{(x)}$.\vspace{2ex}

%The engine is designed to take advantage of the pattern generated by $\mathcal{M}$\m including the free energy locked up in correlations.  See Supplementary Material \ref{app:FreeEnergyDecomp} for a relevant decomposition \MG{Discussion?} of the free energy.
%\vspace{2ex}

%\noindent\emph{Synchronizing to a quantum source}\m

\section{Synchronizing to a quantum source}
\label{sec:synchronize}
To fully leverage the structure of the pattern, the engine must dynamically incorporate
information from past interactions with
the elementary quantum systems, so that the engine's memory becomes correlated with the latent state of the source.
This can be done by tracking\m within the internal memory $M$ of the engine\m an interaction\hyp induced \emph{belief state} $\gbm \eta_t$
about the latent state of the source.

The type of interaction between engine
and fuel at each time can depend on the 
memory of the engine.
The general interaction and observation at time $t$ can be described by a 
positive operator\n valued measure
(POVM)
on the Hilbert space of the elementary fuel system $A_t$;
Let $\Obs_t$ denote the random variable for the observed outcome thereof. 
%of POVM performed on the quantum state at time $t$.

%To fully leverage the structure of the pattern, the engine must implicitly synchronize (to the extent possible) to the latent states of the generator \MG{We never specified what is the generator}
%over time via interactions with the quantum states \MG{readers may not be clear what synchronize means}. 
%
%This can be done by tracking %a finite set of
%the
%\emph{state of knowledge} $\gbm \eta_t$
%about the generator state \MG{first mention of generator state here, not sure what it refers to},
%within internal memory $M$ of the engine. 
The optimal belief state $\gbm\eta_t$ at time $t$ is an observation\hyp induced probability distribution over the latent states $\SSet$ of the source with probability elements
\begin{align}
\gbm\eta_t(s) =\Pr(\St_t=s | \Obs_1 \dots \Obs_t = \littleObs_1 \dots \littleObs_t) ~.
\end{align}
This is the best knowledge that a local classical memory can have, as it represents the actual distribution over latent states as one would calculate via Bayes rule.
It is convenient to treat $\gbm\eta_t$
as a length-$|\SSet|$ row vector for linear algebraic manipulation.
Given a sequence of observations,
what is the probability $P_t(x)$ that 
the source will 
next produce quantum state $\sigma^{(x)}$?
It is 
$P_t(x) = \gbm\eta_t  T^{(x)}  \mathbf{1}$,
where
$\mathbf{1}$ is the column vector of all ones. 
The belief state $\gbm\eta_t$
thus
determines an expectation %prediction
of the next quantum state 
\begin{align}
\xi_t = \sum_{x \in \mathcal{X}} P_t(x) \, \sigma^{(x)} ~,
\label{eq:expected_state}
\end{align}
with more free energy than the local reduced state $\xi_0$ of 
$\rho_{\overleftrightarrow A}$.
%\Rev{provide short proof}.
%\\
This memory enhancement to free energy is proven in App.~\ref{sec:MemoryEnhancementToFreeEnergy}.

%In the asymptotic limit where $t\to\infty$, the local reduced state of $\rho_{\overleftrightarrow A}$ will simply be a probability mixture of $\sigma^{(x)}$ in the form of $\sum_{x \in \mathcal{X}} \gbm\pi  T^{(x)}  \mathbf{1} \, \sigma^{(x)}$. Since $\xi_t$ is formed condition on past observations, the mixture will be less entropic. 

%\MG{Can confuse readers in whether $\gbm\eta_t$ itself is vector, or if it is the $t^{th}$ component of a vector}.
%The state of knowledge $\gbm\eta_t$
%enables a refinement
%of the expected quantum state $\xi_t$ \MG{what does refinement mean in this context?}, with more free energy than the local reduced state of 
%$\rho_{\overleftrightarrow A}$.
%since the former furthermore incorporates knowable correlations.
%In particular,
%\begin{align}
%\xi_t = \sum_{x \in \mathcal{X}} \gbm\eta_t  T^{(x)}  \mathbf{1} \, \sigma^{(x)} ~,
%\end{align}
%where $\mathbf{1}$ is the column vector of all ones \MG{this equation is a bit confusing as it is not clear which indices are multiplying which?}.

The transition rules between these belief states  
are determined via Bayesian inference, based on the anticipated distribution of the observable $\Obs_t$.
If the source is known, but no observations have yet been made, then the optimal belief state is simply the stationary distribution over source states:
$\gbm\eta_0 = \gbm\pi $, 
which satisfies $\gbm\pi = \gbm\pi \sum_{x \in \mathcal{X}}T^{(x)}$.
\vspace{0.5em}
\\ \noindent
\textbf{Theorem 1.}
\emph{For any POVM
on the quantum state of the system at time $t$,
the optimal belief state\m about the 
latent state 
of the quantum source\m updates
iteratively
according to}

\noindent
\scalebox{0.89}{
\begin{minipage}{1.12\columnwidth}
\begin{align}
\gbm\eta_{t} = 
z_t^{-1} \!
\sum_{x \in \mathcal{X} } \!
\Pr( \Obs_t = \littleObs_t | X_t = x, K_{t-1} = \gbm\eta_{t-1} ) \, \gbm\eta_{t-1} T^{(x)}
\label{eq:update}
\end{align}
\end{minipage}
}
\vspace{0.5em}

\noindent
\emph{where $K_t$ is 
the random variable for the belief state,
and 
$z_t = 
\sum_{x' \in \mathcal{X} }
\Pr( \Obs_t = \littleObs_t | X_t = x', K_{t-1} = \gbm\eta_{t-1} ) \, \gbm\eta_{t-1} T^{(x')} \mathbf{1}$
is a normalizing factor.
}
\vspace{0.5em}

We derive Thm.~1 in
Appendix %Supplemental Material 
\ref{app:MSP}, 
%via a generalization of 
generalizing
the so-called mixed\hyp state presentation \cite{crutchfield1994calculi,ellison2009prediction,marzen17nearly, riechers2018spectral, jurgens21shannon}.

The probability
$\Pr( \Obs_t = \littleObs_t | X_t = x, K_{t-1} = \gbm\eta_{t-1} )$
appearing in 
Eq.~\eqref{eq:update}
is typically a straightforward physics calculation
of the probability that the observed POVM outcome $\littleObs_t$ should be obtained, given
that the system was
prepared as $\sigma^{(x)}$.
Conditioning on
the previous state of knowledge $K_{t-1}$
%in Eq.~\eqref{eq:update}
is important 
to the extent that 
it can influence the 
choice of POVM applied at time $t$.

%For any sequence of POVMs on the quantum states,
%the optimal state of knowledge will be updated after every interaction according to 
%\begin{align}
%\label{eq:update_pc1}
%\gbm\eta_{t+1} &=\sum_{x\in \mathcal{X}}
%\text{Pr}(X_{t+1}=x|W_{t+1} = w, \St_t \sim \gbm\eta_t)
%\frac{\gbm\eta_t T^{(x)}}{\gbm\eta_t T^{(x)}\mathbf{1}} ~.
%\end{align}

Notice that, for each $\littleObs$, the update rule for the belief state can be
interpreted as a nonlinear return map. 
If the return map enables prediction, then we have some access
to the nonequilibrium free energy in correlations.
\vspace{0.5em}
\\ \noindent
\textbf{Observation 1.}
\emph{Memory can be thermodynamically advantageous when
the belief-state update rule has an attractor other than the stationary fixed point $\gbm\pi$
for at least one observable outcome.} 

%\\ \noindent
%In the following, we  find that there will generally be phase transitions in the parameter space of processes
%where the return map
%suddenly changes stability to gain a nontrivial attractor 
%and thus knowledge suddenly becomes beneficial. \MG{I am not sure if this paragraph will be meaningful to readers at this point}

%In the present case, we observe work-extraction values $W_t = \Obs_t$ via changes in the energy of the battery. 
%Accordingly,
%the return map
%can be calculated from the physics of the work-extraction protocol. \MG{Nor will this paragraph. $W_t$ not yet introduced!}

%\MG{Section note: Overall I am concerned that the above section will not make much sense to readers. In particular, we have never explain the big picture of what our engine does or is allowed to do. i.e., The engine only has access to the $t^{th}$ qubit at time $t$, it can measure in basis $O_t$, and tracks a belief state... I see a lot of this is is explain in the subsequent section. So there likely needs to be some rearranging of content.}

%\noindent\textbf
%\vspace{2ex}

%\noindent\emph{Engine construction}\m 

\section{Engine construction}
\label{sec:construct}

%\VN{Was the next paragraph originally before this one? It seems we’re introducing the battery there but talking about it above.}

%, and thus updating the engine's corresponding memory state.

The engine is equipped with an internal classical memory $M$ and access to a heat reservoir $R$ at some fixed temperature $T$. Its objective is to extract work by raising the internal energy of a
%an ideal 
work reservoir or `battery' $B$. 
To accomplish this,
it can bring each system $A_t$
closer to its equilibrium state 
$\gamma = e^{- H/ \kB T}/Z$,
%where $H$ is the Hamiltonian that all $\{A_t\}_t$ are restricted to,
%of the elementary fuel system,
$\kB$ is Boltzmann's constant, and 
$Z = \tr(e^{- H/ \kB T})$ is 
%the system's 
the associated
equilibrium partition function that yields the equilibrium free energy $F = -\kB T \ln Z$.
For simplicity of presentation, we assume that each subsystem $A_t$ is subject to the same Hamiltonian $H$, but 
our results generalize in an obvious way
if we allow
different Hamiltonians for each subsystem.
The construction of the pattern engine 
then
requires only the description of the HMM $\mathcal{M}$
and the Hamiltonian $H$.

To harvest the free energy locked up in correlations, 
the engine's internal memory should 
somehow become correlated with the latent state of the source during its energy\hyp harvesting operation.
However, directly measuring each quantum system
would 
disturb the state and potentially cost energy.
%cost energy, as well as disturbing its state and thereby depleting free energy.
Rather, 
our engine updates its memory conditioned on the extracted-work value $W_t$
at each time.  The change in the energy of the battery thus serves as the observable $O_t = W_t$
for updating the belief state.

At each time step, conditioned on its memory state, the engine performs a \emph{work\hyp extraction protocol}---a unitary transformation of
the composite $A_t$, $B$, and
$R$
supersystem, designed 
to transfer energy from $A_t$ to $B$.

%a composite system\m composed of 
%system $A_t$, ideal work-reservoir $B$, and 
%temperature-$T$ thermal reservoirs 
%$A_t$, $B$, and
%$R$\m designed to transfer energy from $A_t$ to $B$.
%As typical in thermodynamics, an \emph{ideal work reservoir} 
%incurs negligible change in its entropy throughout the protocol. 
%;
%as a consequence the reduced dynamics of system $A_t$ and thermal reservoirs $R$ remains unitary. 
%~\footnote{A typical example of a work reservoir 
%would be the position of a massive object, perhaps on a spring or in a gravitational field; it's position determines the Hamiltonian for the system and thermal reservoirs.}  

At best,
the work\hyp extraction protocol at time step $t$
would extract all the
nonequilibrium addition to free energy
$\kB T \text{D}[\rho_t^{*} \| \gamma]$
when it acts on a chosen state $\rho_t^*$,
where 
$\text{D}[\rho \| \gamma] = \tr(\rho \ln \rho) - \tr(\rho \ln \gamma)$ is the quantum relative entropy.
These are the 
$\rho^*$-\emph{ideal work-extraction protocols}, which we define as
any work-extraction protocol
that satisfies the following:
\begin{enumerate}[label=(\roman*)]
    \item When the initial state of the system is $\rho^*$, it achieves %on-average 
zero entropy production,
transferring on-average all nonequilibrium addition to free energy $\kB T \text{D}[ \rho^* \| \gamma]$
to a work reservoir, $B$;
\item It conserves energy globally among 
the reduced states of 
$A_t$, $B$, and $R$
when  acting on 
any eigenstate of $\rho^*$.
\end{enumerate}
Our next theorem fully characterizes the set of extracted values 
and their probabilities for any such protocol (see Appendix \ref{sec:ProofOfWorkStats} 
for the derivation). 
\vspace{0.5em}
\\
\textbf{Theorem 2.}
\emph{Each $\rho^*$-ideal work-extraction protocol
	exhibits at most $d$ distinct extracted-work values.
These extracted-work values can be expressed,
in terms of the ideal input's spectral decomposition
$\rho^* = \sum_{n} \lambda_n \ket{\lambda_n} \bra{\lambda_n}$,
as
}
%	Measured energy change of the work reservoir, 
%	for any}
%$\rho^*$-\emph{ideal work-extraction protocol\m with
%spectral decomposition of the ideal input 
%$\rho^* = \sum_{\lambda} \lambda \ket{\lambda} \bra{\lambda}$\m will have discrete values}
\begin{align}
	w^{(n)} \coloneqq \bra{\lambda_n} H \ket{ \lambda_n}  + \kB T \ln \lambda_n - F 
	\label{eq:WorkExtractionValues}
\end{align}
\emph{with associated probabilities}
\begin{align}
	\Pr \bigl( W \! = w^{(n)}  | \sigma \bigr)
	&= \sum_{m} \! \bra{\lambda_m }  \sigma \ket{ \lambda_m}  \delta_{ w^{(n)}   , w^{(m)}  } 
	\label{eq:WorkExtractionProbs}
\end{align}
 \emph{This set of values is independent of the actual $d$-dimensional quantum state $\sigma$ input to the protocol, although the input state determines the probabilities of each outcome.} 
 %\MG{Does each $\ket{\lambda}$ exactly have eigenvalue $\lambda$?}
%\emph{when the protocol acts on any input-state} $\sigma$.
%
%
%The work-extraction protocol at time-step $t$
%can be
%designed 
%such that the expectation value
%of extracted work is equal
%to the 
%nonequilibrium addition to free energy $\kB T \text{D}[\rho_t^{*} \| \gamma]$
%when acting on state $\rho_t^{*}$,
%where 
%$\text{D}[\rho \| \gamma] = \tr(\rho \ln \rho) - \tr(\rho \ln \gamma)$ is the quantum relative entropy~\cite{SSP14}.
%The spectral decomposition of 
%$\rho_t^* = \sum_{\lambda}
%\lambda \ket{\lambda} \bra{\lambda}$
%plays an important role in the work distribution.
%We find that\m independent of the actual state transformed by the protocol\m there will be up-to $d$ distinct work-extraction values 
%\begin{align}
%    \wlambda = \bra{\lambda} H \ket{\lambda} + \kB T \ln \lambda - F
%\end{align}
%at time $t$
%if $A_t$ is a $d$-dimensional qudit.
%The local input-state $\sigma$
%transformed by the work-extraction protocol 
%determines the probability
%distribution over these
%work-extraction values:
%\begin{align}
%    \Pr(W = \wlambda | \sigma)
%    = \sum_{\lambda'} \delta_{\wlambda, \wlambda[\lambda'] } \bra{\lambda'} \sigma \ket{\lambda'} ~.
%\end{align}
%In particular, this gives 
%

\vspace{0.5em}

Notably, this yields the probability distribution for work extracted when the protocol optimized for $\rho_t^{*}$ actually operates on $\sigma^{(x)}$.
Regardless of how the belief state 
influences the choice of $\rho_t^*$,
we can now leverage Thms.~1 and 2 to rewrite the belief update as
\begin{align}
\label{eq:update_pc2}
\gbm\eta_{t+1} &=
\frac{\sum_{x \in \mathcal{X} }
	\sum_n \delta_{ w_{t+1}, \wlambda[n] } \bra{\lambda_n}  \sigma^{(x)} \ket{ \lambda_n} \, 
	\gbm\eta_{t} T^{(x)} }{
	\sum_n \delta_{ w_{t+1},  \wlambda[n]  } 
	 \bra{\lambda_n} \xi_{t} \ket{\lambda_n} }
~.
\end{align}
With Eq.~\eqref{eq:update_pc2},
the belief-state return maps
now reflect
%are now calculated from 
the physics of the work-extraction protocol.

From Eqs.~\eqref{eq:expected_state} and \eqref{eq:WorkExtractionProbs}, 
we find that the work-induced transitions between belief states have probabilities
\begin{align}
%\Pr(W_{t+1} = \wlambda[n] | K_t = \gbm\eta_t )
%= \sum_m \braket{ \lambda_m | \xi_t | \lambda_m }
%\delta_{\wlambda[n], \wlambda[m]} ~.
%
\Pr(W_{t+1} = w | K_t = \gbm\eta_t )
= \sum_n \braket{ \lambda_n | \xi_t | \lambda_n }
\delta_{w, \wlambda[n]} ~.
\label{eq:BeliefStateTransitionProbs}
\end{align}

Finally, to take thermodynamic advantage of this knowledge,
the work-extraction protocol at each step\footnote{Except at the first step, where we use some $\rho^* \neq \xi_0$, to avoid an unstable fixed point in the knowledge update.} 
is optimized for the expected state, so that $\rho_t^* = \xi_t$.
Indeed, extracting all work from the expected state $\xi_t$
requires a protocol 
designed around this expectation~\cite{riechers2021initial}.
In this case,
the denominator in Eq.~\eqref{eq:update_pc2}
simplifies to 
$\sum_n \lambda_n \, \delta_{ w_{t+1},  \wlambda[n] } $.
Similarly, Eq.~\eqref{eq:BeliefStateTransitionProbs}
simplifies to 
$\sum_n \lambda_n \, \delta_{ w,  \wlambda[n]  } $.

%Details of the derivations can be found in Supplementary Material \ref{app:expected work}.

%Spectral decomp

%We find that 
%work values, and probs

%Update

%be thermodynamically optimal for 
%some state $\rho_t^{*}$,
%such that the expectation value of work extracted would be $\kB T \text{D}[]$

%extract all 
%nonequilibrium free energy 

%of the nonequilibrium addition to free energy, $$,

%\begin{equation}
%\label{eq:update_pc}
%\begin{split}
%    \gbm\eta_{t+1} &=\sum_{x\in \mathcal{X}}\text{Pr}(X_{t+1}=x|W_{t+1} = w_{t+1}, \St_t \sim \gbm\eta_t)\frac{\gbm\eta_t T^{(x)}}{\gbm\eta_t T^{(x)}\mathbf{1}}\\
%    &= \frac{\sum_{x \in \mathcal{X} }
%	\sum_m \delta_{ w, W_{  \lambda_m^{(\gbm\eta_{t}) }}} \bra{\lambda_m^{(\gbm\eta_{t})}}   \sigma^{(x)} \ket{ \lambda_m^{(\gbm\eta_{t})} } \, 
%	\gbm\eta_{t} T^{(x)} }{
%	\sum_m \delta_{ w,  W_{  \lambda_m^{(\gbm\eta_{t})} }    } 
%	 \lambda_m^{(\gbm\eta_{t})} 
%	}~,
%\end{split}
%\end{equation}
%where $\lambda_m$ are the spectrum of the expected state $\xi_t$. Detailed derivation can be found in Supplementary Material \ref{app:expected work}.\\

Combining Eqs.~\eqref{eq:WorkExtractionValues} and \eqref{eq:BeliefStateTransitionProbs},
we compute 
\begin{align}
\langle W_t \rangle =\sum_{\gbm\eta, w} 
w
\Pr(K_{t-1} \! = \gbm\eta) \Pr(W_t = w | K_{t-1} \! = \gbm\eta)
\end{align}
and find
\begin{align}
\label{eq:xi_t_expectation}
    \langle W_t \rangle =
    \kB T \langle \text{D}[\xi_t \| \gamma]  \rangle_{\Pr(K_t)} ~.
\end{align}
Note that the expectation value on the right-hand side is taken over the instantaneous distribution over belief states.
The %resultant 
meta-dynamic over
belief states thus
determines both the transient and asymptotic work-extraction rate.
In particular, the
stationary distribution
over recurrent belief states 
allows closed-form expressions for the asymptotic work-extraction rate. 

%Note that the average on the right-hand side is taken over the distribution of belief states.
%, with the possible work values and probabilities shown in Eq.~\ref{eq:WorkExtractionValues} and \ref{eq:WorkExtractionProbs}.

%are guaranteed to achieve optimal predictive power at minimal memory cost for the underlying stochastic process conditioned on the work statistics of the specific process \cite{ellison2009prediction,crutchfield1994calculi}. %The transition structure among these optimal states, in turn, determines the optimal transition rules for our engine.

\begin{figure} % schematic diagram of the flow
\begin{center}
\includegraphics[width=\columnwidth]{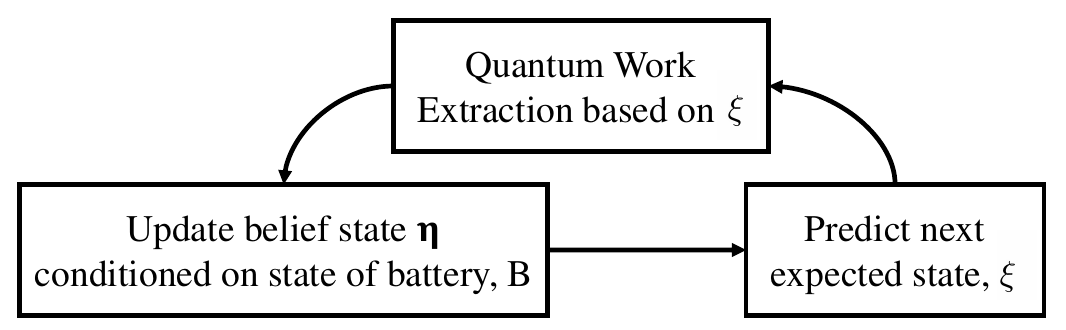}
\caption{The protocol proceeds cyclically to fine-tune the belief state.}
\label{fig:schem_flow}
\end{center}
\end{figure}

Both the belief states and transitions between them
derive from
%can be derived 
%from just 
%the full description of 
the HMM of the
known source.
%underlying HMM.
%, $\mathcal{M}$. 
Belief states can thus be explicitly represented in the memory $M$
of an autonomous work-harvesting device.
%which 
The memory states $\{ (\gbm\eta , \varepsilon) \}_{\gbm\eta, \varepsilon}$
should also store the last measured energy state $\varepsilon$ of the battery.
A memory-controlled unitary
can implement
memory-assisted quantum
work extraction, as depicted in the circuit diagram of Fig.~\ref{fig:schem_eng}.
Subsequent measurement of the battery state then 
gives access to the work extracted and allows an autonomous update of the memory,
according to the above-outlined rules of Bayesian prediction.
This prediction\n extraction cycle continues repeatedly, 
as suggested in Fig.~\ref{fig:schem_flow}.

In short, at every time step, the engine performs both prediction and work-extraction subroutines.
The prediction subroutine updates the 
belief state stored in $M$ via pre\hyp programmed transition rules conditioned on the energy state of $B$. 
Each unique $\gbm \eta_t$---up to the desired memory resolution---corresponds to a subset of distinguishable states in the memory. 
%and implicitly corresponds to an expected quantum state.
The engine's work-extraction subroutine is conditioned on the memory $M$ and acts on $A_t$ with a quantum work-extraction protocol 
%of \cite{SSP14} (presented in detail in Supplementary Material~\ref{app:WEprot}),
tailored to extract work from $\xi_t$.

%We effectively shifts the dynamic between the latent states $\SSet$ that is conditioned on the emission of quantum states, which cannot be directly observed, $(\sigma^{(x)})_x$ to the meta-dynamic between $(\gbm \eta_t)_t$ that is conditioned on physical observables.

%\noindent \textbf{Functioning}--As shown in Supplementary Material \ref{app:FreeEnergyDecomp},
%the free energy in the quantum process\n fuel lies in the interplay of two aspects: (1) quantum states of individual elementary systems and (2) classical temporal correlations. The engine's functioning address these two aspects through specialized techniques as depicted in Fig \ref{fig:schem_flow}. The more detailed functioning algorithm of the engine can be found in Box~\ref{algorithm} 

\section{Example processes and alternative approaches} 
\label{sec:example}
To demonstrate our \emph{memory-assisted quantum} approach\m using a quantum work-extraction
protocol designed for the work\hyp observation\n induced 
expected state $\rho_t^* = \xi_t$\m we apply it to the quantum perturbed-coin and golden-mean processes depicted in Fig.~\ref{fig:perturb_coin}. We compute our engine's long-term work output from 
%these processes 
this approach (i)
and compare its performance with those of three alternative approaches (ii - iv):
\begin{enumerate}[label=(\roman*)]
    \item 
    \label{approach:1}
    \emph{memory-assisted quantum}, where the quantum work-extraction
protocol is optimized for the work\hyp observation\n induced 
expected state $\rho_t^* = \xi_t$;
    \item 
    \label{approach:2}
    \emph{memory-assisted classical},
where the protocol, unable to extract work from quantum coherences,
is optimized for the energy-dephased version of the expected state $\rho_t^* = 
\xi_t^{\text{dec}} \coloneqq \sum_E \ket{E}\bra{E} \bra{E} \xi_t \ket{E}$;
%instead of unitary rotation mentioned in step 2-4 of Box \ref{algorithm}, one does a projection onto Hamiltonian basis instead
\item 
\label{approach:3}
\emph{memoryless quantum processing}, 
where memory is never updated by observations, and the quantum work-extraction protocol
is simply optimized for the time-averaged quantum state
$\rho^* = \xi_0 = \sum_{x} \gbm\pi  T^{(x)}  \mathbf{1} \, \sigma^{(x)} $; and%where the state of knowledge remains at the stationary distribution of the process $\gbm\eta_0=\gbm\pi$
\item 
\label{approach:4}
\emph{overcommitment to most probable quantum state}, 
%where the protocol is 
with protocol optimized for $\rho_t^* = \sigma^{(\text{argmax}_{x} \gbm\eta_t  T^{(x)}  \mathbf{1} )} $.
\end{enumerate}
%where the protocol can only operate on $\sigma^{(x)}$s.

%{\color{red}(PMR: I don't think (iii) actually describes how Ruo Cheng simulated that part.  (From what I recall, that part of the simulation was not well motivated.)  If not, I suggest we either re-do that part of the simulation to align with the description, or remove the analysis and discussion of (iii) altogether; perhaps relegating to Supp.\ Mat.)}

%For comparison, recall that in our full quantum memory-assisted approach,
%the work-extraction protocol is designed for the expected quantum state
%$\rho_t^* = \xi_t$.
%Also note that the states of knowledge will be different in each approach.
%In fact, there is a significant memory advantage in the quantum-memory case,
%where there are only two recurrent states, rather than the infinite number of recurrent memory states in the less energetically rewarding classical processing case.

%\clearpage

\begin{widetext}

%\begin{figure*}[!htbp]
\begin{figure*}[t]
\begin{center}
\includegraphics[width=0.9\textwidth]{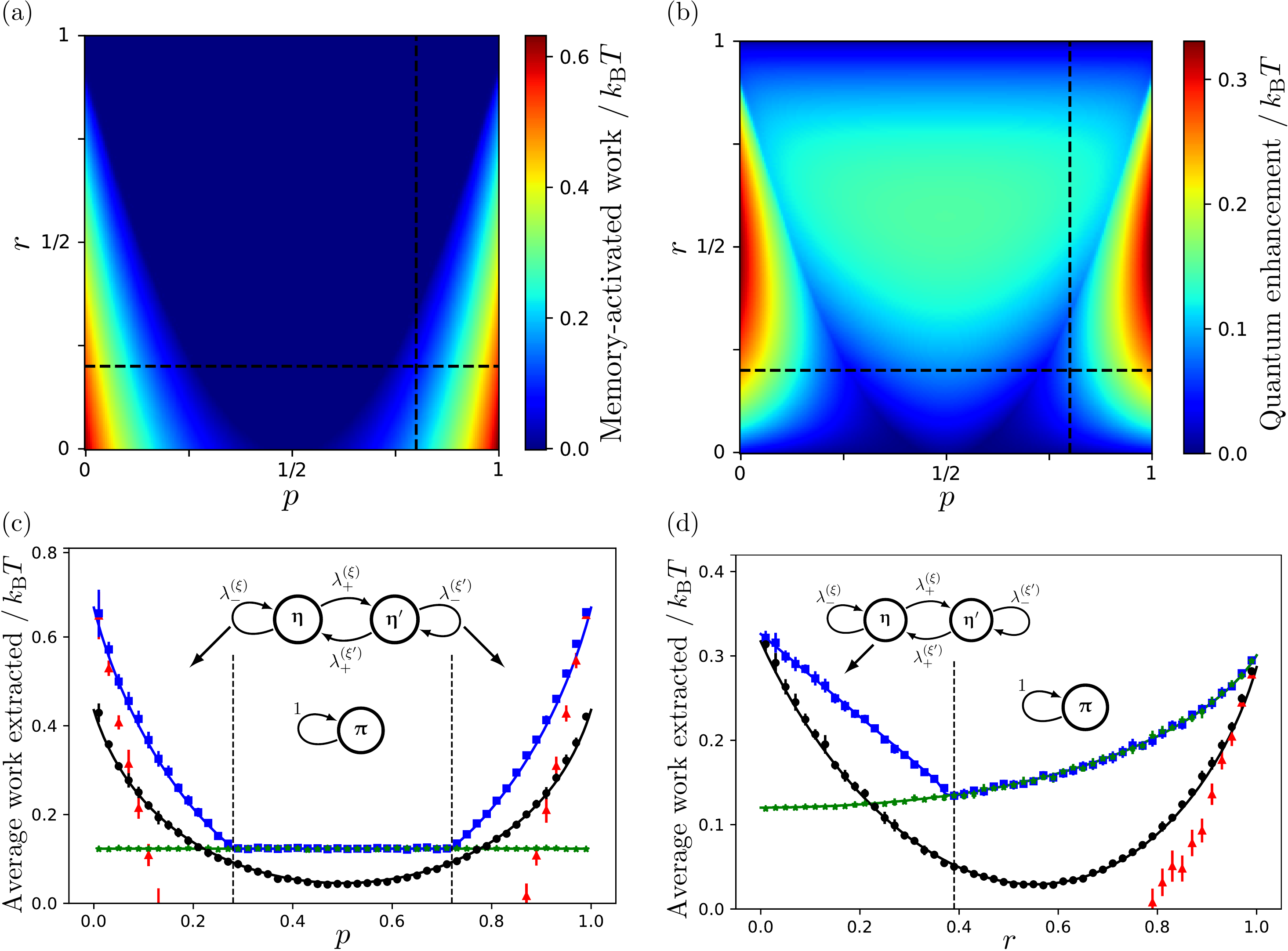}
\caption{Comparison between average work-extraction rates of various approaches.
%Recall that 
$p$ characterizes the transition probability between the two latent states of the perturbed-coin process, and $r$ quantifies the overlap between the two quantum outputs. (a) Memory enhancement of work extraction. 
(b) Quantum enhancement of work extraction. Panels (c) and (d) reveal phase transitions in memory enhancement through cross-sections of parameter space. Analytic results (solid lines) and  simulations (markers) are shown. Blue (squares) represents approach \ref{approach:1}; black (circles) represents approach \ref{approach:2}; green (stars) represents  approach \ref{approach:3}; red (triangles) represents approach \ref{approach:4}.}
\label{fig:Deg}     
\end{center}
\end{figure*}
%\begin{table*}[htbp]
%    \begin{center}
%    \caption{Summary of metadynamics in different regimes. The update function shows the nonlinear relationship between $\epsilon_t$ and $\epsilon_{t+1}$. The belief evolution shows the evolution of $\epsilon_t$ over iterations, which give rise to the corresponding work series, with two possible work values per belief state. The recurrent belief states show the recurrent metadynamic of the different regimes.}
%    \label{tab:meta}
%\includegraphics[width=1.0\textwidth]{Table_for_QPE_crop.pdf}
%\end{center}
%\end{table*}

\end{widetext}

The asymptotic work-extraction rates
$\lim_{t \to \infty} \langle W_t \rangle $ from these approaches are compared 
%These approaches are all compared 
in Fig.~\ref{fig:Deg}, 
both analytically and with numerical simulations, 
%\MG{Some repetition with sentence above},
for the perturbed-coin process. 
%of Fig.~\ref{fig:perturb_coin}(a).
In Fig.~\ref{fig:Deg}(a) memory-activated work is defined as the difference between memory-assisted quantum approach \ref{approach:1} and memoryless quantum approach \ref{approach:3}. In Fig.~\ref{fig:Deg}(b) quantum enhancement is defined as the difference between memory-assisted quantum approach \ref{approach:1} and memory-assisted classical approach \ref{approach:2}.\\

In this demonstration,
the system extracts work from a sequence of qubits,
each with Hamiltonian $H = E_0\ket{0}\bra{0}+E_1\ket{1}\bra{1}$,
with $E_1 - E_0 = \kB T$ in this case.
At each time step, 
the source
produces one of two nonorthogonal quantum states:
$\sigma^{(0)}=\ket{0}\bra{0}$ or $\sigma^{(1)}=\ket{\psi}\bra{\psi}$, where $\ket{\psi}=\sqrt{r}\ket{0}+\sqrt{1-r}\ket{1}$, 
according to the labeled transition matrices of 
the perturbed-coin process. 

As seen in Fig.~\ref{fig:Deg}, our memory-assisted quantum approach \ref{approach:1} always performs at least as well as all other approaches, and strictly outperforms them in many regions of parameter space. 
In these examples,
the ideal input $\rho_t^*$
is a qubit density matrix with eigenvalues $\lambda_{\pm}^{(\rho_t^*)}$, where $\lambda_+^{(\rho_t^*)} \geq\lambda_-^{(\rho_t^*)} \geq 0$.
Due to
Eqs.~\eqref{eq:WorkExtractionValues} and \eqref{eq:WorkExtractionProbs}, 
we thus expect
to observe one of two possible work values, $w^{(\pm)}$
from each distinct belief state.

Approaches
\ref{approach:1}-\ref{approach:3}
share some nice features.
From Eq.~\eqref{eq:WorkExtractionProbs}, assuming distinct work values $w^{(+)} \neq w^{(-)}$, we find that the 
probability of observing
each possible work value is simply given by the corresponding eigenvalue of the optimal input:
\begin{align}
\Pr(W_{t+1} = w^{(\pm)} | K_t = \gbm\eta_{t}) = \lambda_{\pm}^{(\rho_t^*)} ~.
\end{align}
%In particular $\dots$
Combining this with Eq.~\eqref{eq:WorkExtractionValues}, we find that 
the rate of work extraction can be expressed as 
\begin{align}
\label{eq:rhostar_expectation}
    \langle W_t \rangle =
    \kB T \langle \text{D}[\rho_t^* \| \gamma]  \rangle_{\Pr(K_t)} 
\end{align}
for approaches \ref{approach:1}-\ref{approach:3}---although, notably, both $\rho_t^*$ and the set of belief states will be different in each approach---with $\rho_t^* = \xi_t$, $\xi_t^\text{dec}$, and $\xi_0$, respectively.
The non-negativity of relative entropy thus guarantees the non-negativity of expected work extraction from these three approaches.
Note that Eq.~\eqref{eq:rhostar_expectation} is more general than Eq.~\eqref{eq:xi_t_expectation} and allows us to 
compare work-extraction performance of the three approaches.

%work out the expectation values for work values when we consider different protocol.
%[By the information processing inequality, this also proves that our memory-assisted quantum approach (i) 
%is always better at extracting work than the memory-assisted classical approach (ii).
%It also strongly suggests that 
%our approach will never be worse than the memoryless quantum processing approach.]

%(These are not features of approach (iv), which suffers extreme drawbacks that we will come back to shortly.)

%, which we label ``plus" and ``minus" work values. 
%Likewise, the probability of measuring these values given a specific belief state $\gbm \eta$ will be denoted by $\lambda_{\pm}^{(\gbm \eta)}$. The d

Without any memory, the 
extractable structure is limited to the
time\hyp averaged statistical bias of the output \cite{Boyd17PRE}, which explains why memoryless work extraction varies with $r$ but not $p$. 
Further details of the analytic solution for expected work extraction can be found in Appendix %Supplemental Material 
\ref{app:expected work}.

%At each time step, the extraction protocol extracts energy $w_t$. 
%The average work extracted per time step, in the long-time 
%The asymptotic work-extraction rate
%$\lim_{t \to \infty} \langle W_t \rangle $ 
%is shown in Fig.~\ref{fig:Deg} for all four approaches.

Our numerical simulations use the quantum work-extraction protocol
of Ref.~\cite{SSP14} %designed for $\rho_t^*$ 
at each time-step, and agree very well with our more general analytical predictions.
Indeed, the quantum work-extraction protocol of Ref.~\cite{SSP14} 
provides an example of a $\rho^*$-ideal work-extraction protocol, 
in the limit of many bath interactions. 
More details can be found in %Supplemental Materials 
Appendices
\ref{supp:WEprot} and \ref{supp:simulation}.

%\noindent\textbf{Simulation}\m The Hamiltonian of the system is set to be $H=E_0\ket{0}\bra{0}+E_1\ket{1}\bra{1}$. A stochastic string of length $n$ consisting of $\sigma^{(0)}=\ket{0}\bra{0}$ and $\sigma^{(1)}=\ket{\psi}\bra{\psi}$ where $\ket{\psi}=\sqrt{r}\ket{0}+\sqrt{1-r}\ket{1}$, is generated using the perturbed coin process. At each time step, the extraction protocol will extract energy, $\varepsilon$. The average work extracted per time step, $\bar{\varepsilon}=\frac{1}{n}\sum_{i=1}^{n}\varepsilon_i$, for all four different approaches, are shown in Fig.\ref{fig:Deg}. More details can be found in Supplementary Material \ref{supp:simulation}.
%do not account for temporal patterns.
%[In case we get data on non-Markovian processes] As expected, the advantage of our engine is more pronounced for processes with higher levels of temporal correlations. [merge with existing discussion.]

It is tempting to commit to the most likely outcome.
However, the overcommitment approach \ref{approach:4}
performs the worst, since any reset operation (to $\gamma$ in this case)
with minimal entropy production for a pure-state input leads to infinite heat dissipation when operating on any other input~\cite{riechers2021initial, riechers2021impossibility}.
This translates to infinite negative work extraction $\braket{W_t} = -\infty$ in this case of 
$\rho_t^* \in \{ \ket{0} \! \bra{0} , \, \ket{\psi} \! \bra{\psi} \}$.
%$\lambda_-^{(\ket{0} \bra{0})} = \lambda_-^{(\ket{\psi} \bra{\psi})} = 0$.
This divergence can alternatively
be seen from 
Eq.~\eqref{eq:WorkExtractionValues}
as $w^{(-)} \sim \ln \lambda_- \to \ln 0 = -\infty$.
In our numerical simulations,
following Ref.~\cite{SSP14},
this minimal eigenvalue $\lambda_-$
is inversely proportional
to the number $N$ of bath interactions,
so that the overcommitment work penalty diverges as $-\kB T (1-r) \min( p, 1-p ) \ln N$.
%consistent with our simulations.

%\vspace{2ex}

%\begin{widetext}

\begin{table*}[htbp]
    \begin{center}
    \caption{Summary of metadynamics in different regimes. The update function shows the nonlinear relationship between $\epsilon_t$ and $\epsilon_{t+1}$. The belief evolution shows the evolution of $\epsilon_t$ over iterations, which give rise to the corresponding work series, with two possible work values per belief state. The recurrent belief states show the recurrent metadynamic of the different regimes.}
    \label{tab:meta}
\includegraphics[width=1.0\textwidth]{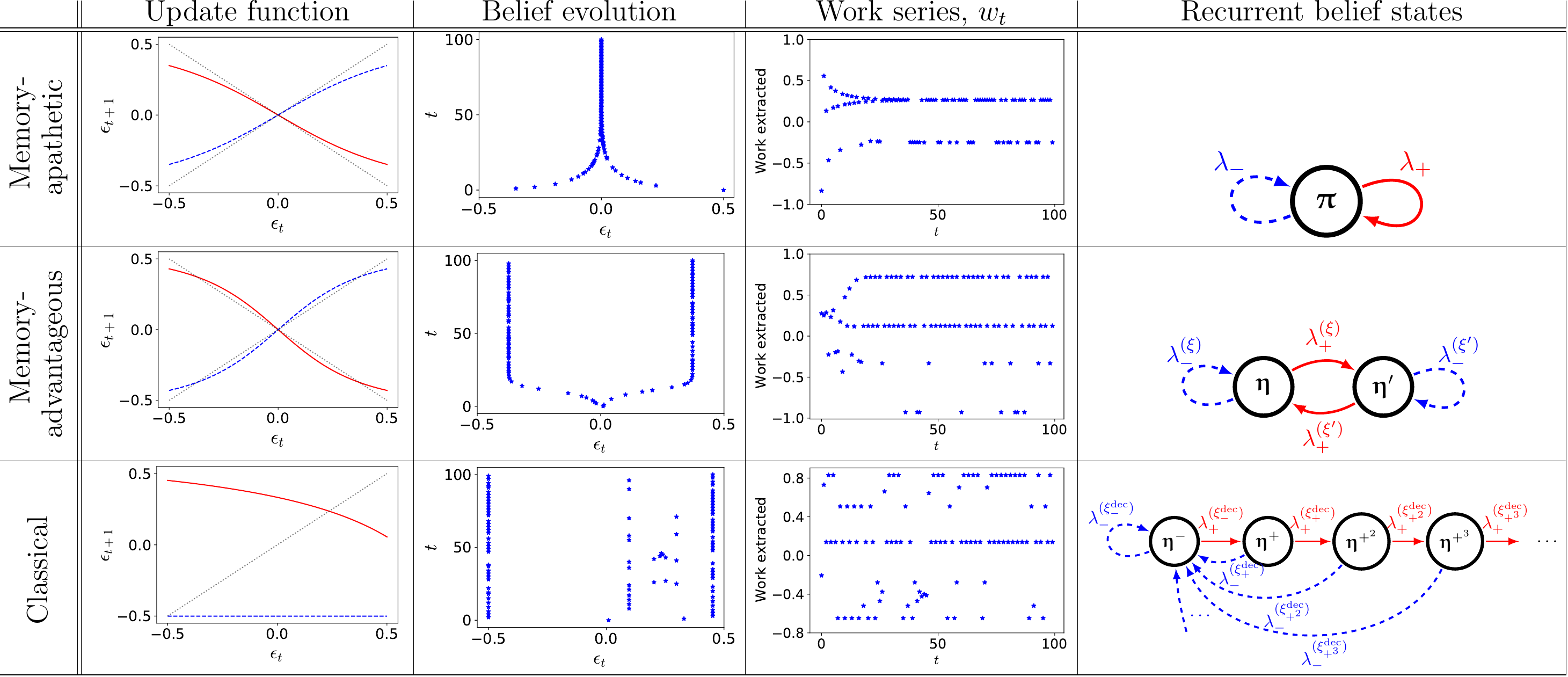}
\end{center}
\end{table*}

%\end{widetext}

%\begin{widetext}

%\begin{table*}[htbp]
%    \begin{center}
%    \caption{Summary of metadynamics in different regimes. The update function shows the nonlinear relationship between $\epsilon_t$ and $\epsilon_{t+1}$. The belief evolution shows the evolution of $\epsilon_t$ over iterations, which give rise to the corresponding work series, with two possible work values per belief state. The recurrent belief states show the recurrent metadynamic of the different regimes.}
%    \label{tab:meta}
%\includegraphics[width=0.95\textwidth]{Table_for_QPE_crop.pdf}
%\end{center}
%\end{table*}

%\end{widetext}

\subsection{Phase transitions in efficacy of knowledge}
Surprisingly, there exists a blue inner region of panel \ref{fig:Deg}(a) where the memoryless quantum approach \ref{approach:3} achieves the same performance
as our memory-assisted quantum approach \ref{approach:1}. 
\emph{There exists a sharp phase boundary within which the use of memory does not boost performance.} 
As seen clearly in panels \ref{fig:Deg}(c) and \ref{fig:Deg}(d), 
the phase boundary exhibits a discontinuity in the first derivative of work extraction with respect to process parametrization.
%Within this region, the process's randomness seemingly overrides the amount of information one gains by observing the past sequence. 
%This causes the engine's inability to synchronize with the internal state of the machine, resulting in a fuzzier state of knowledge over time. 
This phase boundary is not unique to the perturbed-coin process and, indeed, also occurs
in the 2-1 golden-mean process. 

Such phase transitions 
originate from 
bifurcations of the 
attractors
of the belief-state update maps.
This 
is illustrated in
%can be understood from 
Table \ref{tab:meta},
which shows the nonlinear return maps 
%before and after bifurcation, in the first and second row respectively,
along with the consequences for belief dynamics and work-extraction dynamics.
We focus for now on the first two rows of Table \ref{tab:meta},
which illustrate the nonlinear dynamics of our memory-assisted quantum approach \ref{approach:1}, 
in the memory-apathetic and memory-advantageous regimes,
before and after bifurcation respectively.

Recall from Fig.~\ref{fig:perturb_coin}(a)
that 
a two-state machine generates the perturbed-coin process.
Hence,
a scalar
$\epsilon_t \in [-\tfrac{1}{2}, \, \tfrac{1}{2}]$
%fully describes 
suffices to describe
the time-dependent belief state $\gbm\eta_t = (\tfrac{1}{2} +\epsilon_t, \, \tfrac{1}{2} - \epsilon_t)$.
The magnitude of this scalar $\epsilon_t$ 
indicates the 
%confidence
strength of evidence 
%to implicate
that the process is in 
a particular hidden state.

The first column of Table~\ref{tab:meta} 
shows the return maps for $\epsilon_t \mapsto \epsilon_{t+1}$
induced by either 
$w^{(+)}$ (red solid graph) or
$w^{(-)}$ (blue dashed graph),
when the work-extraction protocol is optimized for $\rho_t^* = \xi_t$ (for the first two rows)
or $\rho_t^* = \xi_t^\text{dec}$
(for the last row).
To aid the visual bifurcation and stability analysis, we 
%have drawn 
include
a dotted diagonal line with slope one---representing the identity map---and a dotted diagonal line with slope minus one---representing the swap map.

Intersections between a return map and the dotted identity line
would indicate a fixed point of the map upon its repeated application.
If the magnitude of the slope at the intersection is less than unity, then it is a stable fixed point; 
if the magnitude of the slope at the intersection is greater than unity, then it is an unstable fixed point.
Within the memory-apathetic region of parameter space,
Work extraction does not supply enough evidence to nudge an observer out of a state of complete ignorance. 
In this regime, 
memory does not enhance
quantum work extraction. 
%is not enhanced by memory.

At the phase boundary
in the process' parameter space,
the fixed point at $\gbm\pi$
becomes unstable,
and new attractors emerge for each map.
However, the coexistence of the maps introduces competition between the attractors,
%inducing a non-trivial steady state 
as the maps are
%However, t
%The maps are 
selected stochastically with probabilities $\lambda_{\pm}$.
The two maps (red solid and blue dashed) interact to induce a steady-state metadynamic over recurrent belief states, 
%which is 
shown as a Markov process in each row of Table \ref{tab:meta}'s last column.
In this memory-advantageous region,
work extraction supplies sufficient evidence to 
inform an observer about the hidden state of the process, which in turn avails more extractable work.

The elegance of memory-assisted quantum work extraction 
is reflected in the simple one and two state recurrent memory structures.
In comparison, the classical extractor not only harvests less work, but requires more memory to achieve its relatively meager returns.
Note the infinite number of recurrent memory states in the  classical-processing case,
in the last row of Table \ref{tab:meta}.

%It is interesting that the quantum-memory advantage persists even in the classical limit of $r \to 0$,
%although not when $r=0$ exactly.

%Clearly it is disadvantageous to harvest less work in the classical case, where energetic coherences cannot be harnessed.  However, there is another unexpected disadvantage to classical energy harvesting:
%A significant memory burden, as indicated in 
%the last column of Table \ref{tab:meta}. 

%In fact, the memoryful quantum case achieves significant memory parsimony,
%with only two recurrent states
%in the memory-advantageous region of the perturbed-coin process, 
%compared to the infinite number of recurrent memory states in the less energetically rewarding classical-processing case.

%The classical approach only has access to a pair of orthogonal basis, the act of projection onto said basis decoheres the quantum state and hence the free energy in the coherence cannot be extracted. 

%To benchmark the performance of the memory assisted protocol on process with higher Markov order, 2-1 Golden mean process was chosen. The asymptotic average density matrix of the memoryless approach for both models are kept the same, $\bar{\rho}=\frac{1}{2}(\sigma^{(0)}+\sigma^{(1)})$. 

%\noindent\textbf

%\vspace{2ex}

%\noindent\emph{Thermodynamics of synchronization}\m

%\vspace{2ex}

%\noindent\emph{Transient thermodynamic cost of synchronization}\m 
%of memory update}\m
\section{Fundamental limits of quantum pattern engines}
\label{sec:FunLimits}
\subsection{Initial investment and ideal work-extraction rate}
Has our engine performed optimally?
To answer this requires some benchmark of optimality.
Fortunately, this benchmark is furnished by an appropriate application of the thermodynamic second law.

If we treat the engine's memory 
and the multi-partite quantum pattern 
as a single joint system,
we can apply the second law of thermodynamics to 
(i)
bound
the maximal possible work extraction,
and (ii) discover an initial thermodynamic investment required for the engine to become correlated with the pattern.
The transient work investment is apparent in the `work series' column of Table \ref{tab:meta}.
%This transient energetic cost is 
It is
a necessary price to harvest work at the optimal steady-state rate.

Let us denote the joint state of a $d_\text{M}$-dimensional memory and the length-$L$ pattern as $\rho_t^{(\text{M}, 1:L)}$, with corresponding reduced states of the memory $\rho_t^{(\text{M})}$
and of the pattern $\rho_t^{(1:L)}$.
We will assume that the memory is fully energetically degenerate
so that unitary memory updates do not cost energy.
Then $\boldsymbol{\gamma} = (I/d_\text{M}) \otimes \gamma^{\otimes L}$ describes the equilibrium state of the joint supersystem.
If we denote $W_t^\text{ext} = \sum_{\ell=1}^t W_\ell$ as the net work extracted up to time $t$,
then the second law of thermodynamics 
%can be stated as an 
tells us that the reduction in nonequilibrium free energy 
$\mathcal{F}_t^{(\text{M}, 1:L)}$
upper bounds
the expectation value of extracted work~\cite{parrondo2015thermodynamics}:
\begin{align}
\braket{W_t^\text{ext}} \leq \mathcal{F}_0^{(\text{M}, 1:L)} - \mathcal{F}_t^{(\text{M}, 1:L)} ~.
\end{align}
Since the equilibrium free energy is unchanged during our work-extraction process,
the change in nonequilibrium free energy is proportional to the change in relative entropy between 
$\rho_t^{(\text{M}, 1:L)}$ and  $\boldsymbol{\gamma} $.
This relative entropy simplifies
further when we invoke several features of our framework:
Since the memory is always initialized as $\rho_0^{(\text{M})}$,
the memory and pattern are initially uncorrelated:
$\rho_0^{(\text{M}, 1:L)} = \rho_0^{(\text{M})} \otimes \rho_0^{(1:L)}$.
Through the sequence of deterministic updates, the entropy of the memory is unchanging.
Altogether, these features imply that the extractable work is upper bounded by
\begin{align}
\frac{\braket{W_t^\text{ext}}}{ \kB T} \leq 
\text{D}[ \rho_0^{(1:L)} \| \gamma^{\otimes L}] - 
\text{D}[ \rho_t^{(1:L)} \| \gamma^{\otimes L}]
- \text{I}_t 
%(M; \, 1 \! : \! L) 
~,
\end{align}
where 
$\text{I}_t
%(M; \, 1 \! : \! L) 
= \text{D}[ \rho_t^{(\text{M}, 1:L)} \| \rho_t^{(\text{M})} \otimes \rho_t^{(1:L)}  ]$
is the mutual information that has built up between the memory and the quantum pattern.

Invoking two more features leads us to a more interpretable version of this result.
We note that after $t$ time-steps, the first $t$ subsystems of the pattern have been brought to their equilibrium states, while the other subsystems remain unaltered:
$\rho_t^{(1:L)} = \gamma^{\otimes t} \otimes \rho_0^{(t+1:L)}$.
Let $\text{S}(\rho) = -\tr(\rho \ln \rho)$ denote the von Neumann entropy of $\rho$.
If the engine operates on a very long quantum pattern
with a \emph{von Neumann entropy rate} $\entropyrate \coloneqq \lim_{L \to \infty} \tfrac{1}{L} \text{S}(\rho_0^{(1:L)})$,
%this leads to 
we find
that the steady-state work-extraction rate
is upper bounded by 
%the sum of the local nonequilibrium addition to free energy $\text{D}[\rho_0^{(\ell)} \| \gamma]$
%
%Dividing my $t$ yields an upper bound for the asymptotic rate
\begin{align}
\frac{\braket{W_t}}{\kB T} \, \leq \,
 \text{S} (\rho_0^{(\ell)}) - \entropyrate  +  \text{D} [\rho_0^{(\ell)} \| \gamma] \, \eqqcolon \, \frac{\wideal}{\kB T} ~,
\end{align}
where the choice of $\ell \in \{ 1, 2, \dots , L\}$ is arbitrary since we assume a stationary quantum pattern.
This is a special case of the 
Quantum Information Processing Second Law (QIPSL), derived 
as Eq.~\eqref{eq:QIPSL}
in Appendix \ref{app:QIPSL}.

%$\lim_{L \to \infty}
%\tfrac{1}{L} \braket{W_L^\text{ext}}$

Whenever the input pattern is far from being fully consumed, 
%such that $L-t$ is sufficiently large,
we find that
$\braket{W_t^\text{ext}} \leq t  \wideal - \kB T \, \text{I}_t $.
By the information-processing inequality, the investment cost $\text{I}_t$ is no more than the \emph{quantum excess entropy} $\mathcal{E} 
= \lim_{\ell \to \infty} \text{I} (\rho_0^{(1:\ell)} ; \rho_0^{(\ell+1:2 \ell)})$,
%$ \coloneqq \lim_{L \to \infty} [ \text{S}(\rho_0^{(1:L)}) - L \entropyrate ]$,
which is the quantum mutual information between the sequences of past and future inputs.
However, the engine will only be able to harvest work at the ideal steady-state rate 
if the memory stores all of the information from past inputs relevant for predicting future inputs, such that $\text{I}_t = \mathcal{E}$.
Only then will the process be seen at the true entropy rate $\entropyrate$; otherwise it will appear more random than it really is, with the hidden structure evading extraction.

On the other hand, when 
the entire pattern has been consumed, 
$\text{I}_L = 0$ yet
\begin{align}
\braket{W_L^\text{ext}} \leq L \wideal - \kB T \mathcal{E} ~.
\end{align}
The quantum excess entropy $\mathcal{E}$ is a thermodynamic investment for any ideal extractor---even one that operates non-locally.
However, for local extractors, this investment must be payed upfront as $\text{I}_t$.

%Likely,
%It seems plausible that

In the current framework,
it remains an open question whether 
the knowledge about future inputs $\text{I}_t$
attains %would be less than 
the fundamental limit of what is knowable $\mathcal{E}$,
and 
whether
the work extraction rate $\braket{W_t}$ 
%would be strictly less than 
attains 
$\wideal$.
Perhaps
quantum discord prevents local extractors from performing optimally unless they are equipped with quantum memory.
Unfortunately, there is no known closed-form solution for the von Neumann entropy rate $\entropyrate$, even for the simple latent-state models studied here; Accordingly,
we cannot analytically 
evaluate 
how close our engine is to optimal.
Future studies could elucidate the path to optimality, by investigating the work-extraction rate as 
the agent 
(a)
operates on longer subsequences
or
(b)
explores non-greedy harvesting strategies.

%The net work extraction is penalized by an initialization fee of at least $\kB T \text{I}_t$

%\vspace{2ex}

\subsection{Energetically-free memory updates in steady state}
The initial %transient dissipation 
thermodynamic investment $\text{I}_t$
can be associated with the initial nonunitarity of the work-conditioned map on the memory space---the state compression needed for an observer to synchronize (to the extent possible) with the hidden state of the process.
However, in the steady state,
the work-conditioned maps are unitary for our memory-assisted quantum approach \ref{approach:1}.
This can be seen in the first two rows of the last column of Table \ref{tab:meta}:
For the perturbed-coin process,
the work-induced map is either the identity or swap map on the
restriction to the recurrent belief states.  
These asymptotic memory-update maps are unitary and so can be performed with zero dissipation.\footnote{In contrast, for the classical extractor,
the non-unitarity of the steady-state map induced by $w^{(-)}$ seems to reflect the loss of free energy upon decoherence in the energy basis.}

Further investigation is required to determine if the steady-state work-conditioned maps of our approach are always asymptotically unitary
for other processes.
If not,
it may be beneficial to 
merge 
the work-extraction and prediction subroutines into a single more sophisticated asymptotically-unitary transformation. 
We anticipate that 
the memory updates 
would then be 
energetically free in the steady state.
This expectation aligns with previous related studies on local work-harvesting pattern extractors in the classical domain,
where the ideal pattern extractor---which must be predictive of its input---can update its memory for zero energy cost during  steady state work extraction~\cite{Boyd17Transient, garner2017thermodynamics, Boyd18Thermodynamics, Garner21Fundamental}.

%The classical approach only has access to a pair of orthogonal basis, the act of projection onto said basis decoheres the quantum state and hence the free energy in the coherence cannot be extracted. 

%To benchmark the performance of the memory assisted protocol on process with higher Markov order, 2-1 Golden mean process was chosen. The asymptotic average density matrix of the memoryless approach for both models are kept the same, $\bar{\rho}=\frac{1}{2}(\sigma^{(0)}+\sigma^{(1)})$. 

%\noindent\textbf
%\vspace{2ex}

\section{Discussion}
\label{sec:discussion}
We developed the theoretical prototype for a \emph{quantum pattern engine}: a machine that can adaptively extract useful work from quantum stochastic processes by exploiting knowledge of the temporal patterns they contain. 
We witnessed that, in the presence of coherence, the memory-assisted quantum approach will always outperform the memory-assisted classical approach.
We also demonstrated its advantage over engines that can
only harness static quantum resources\m although, surprisingly,
we found a phase transition marking the onset of memory advantage.
It is an open question whether this 
phase transition coincides with the onset of quantum discord.

%\VN{Should we implicate coherence or discord?}

%or dynamical classical resources (i.e., temporally\hyp correlated \emph{classical} systems) alone. 
%We also witnessed its superiority to adaptive engines that bet on the most likely outcome. 
%The analytical calculation also showed that in the presence of coherence, the memory-assisted quantum approach will always outperform the memory-assisted classical approach.

In Thm.~1, 
we found how to update the 
state of knowledge about any latent-state generator of a quantum process, given any POVM on the current quantum output.
In Thm.~2,
we found the exact work distribution obtained from any $\rho^*$-ideal work-extraction protocol operating on any quantum state. 
This enabled belief-state updates via observed values of work extraction.
Sec.~\ref{sec:FunLimits}
developed the fundamental thermodynamic limits of work extraction from correlated multi-partite quantum systems,
which sets the ultimate benchmark for our approach 
or any alternative.

Despite the advances presented here, many open questions remain for future work.

It is known that measurements themselves can cost some resource, energetic or otherwise~\cite{Guryanova2020Ideal, Elouard2017Extracting}.
Careful accounting of the resources required for (perfect or imperfect) measurement in our framework would provide a fuller picture of our engine's performance.
Alternatively, it
may be possible to 
use a quantum memory and fully unitary evolution (rather than projective measurements on the battery and subsequent conditional operations on the memory), which would avoid this issue completely.

Although designing the protocol for the observation-induced expected state, $\rho_t^*=\xi_t$,
guarantees maximal work extraction
locally in time,
it remains an interesting open question whether
there is a superior steady-state approach that 
sacrifices short-term work extraction 
for greater knowledge and long-term returns.
%On its surface, this seems like the familiar artificial-intelligence tradeoff between exploration and exploitation; 
%however, with a known model, the exploration is more intimidatingly one in time-extended protocol space.

%The energetic cost of measuring the battery's energy and implementing the memory update requires further investigation. 
%This will be an important loophole to close if one desires a truly autonomous energy harvester.
%We anticipate that  
%the energetic cost to update memory may vanish when the work-extraction and prediction subroutines are merged into a single more sophisticated unitary transformation. 

It may be possible to extend our method to more complex quantum processes, e.g.,\ to those with entangled temporal correlations. This would, however, 
likely require a quantum memory.
%for the purpose of entanglement swapping.
%call for a development of predictive modeling techniques beyond the current state of the art.
On the other hand, our method can immediately be adapted to applications where the pattern is spatial instead of temporal (e.g., states of many\hyp body systems), and where the engine is constrained to operate locally on small regions at a time.

%\VN{I would remove
%“for the purpose of entanglement swapping”: it is too
%specific an operation. We could use a quantum memory
%in any number of other ways.}
%\vspace{2ex}

\section{Acknowledgements}
We acknowledge the support of the Singapore Ministry of Education Tier 1 Grants RG146/20 and RG77/22, the NRF2021-QEP2-02-P06 from the Singapore Research Foundation and the Singapore Ministry of Education Tier 2 Grant T2EP50221-0014, the Agency for Science, Technology and Research (A*STAR) under
its QEP2.0 programme (NRF2021-QEP2-02-P06) and and the FQXi R-710-000-146-720 Grant “Are quantum agents more energetically efficient at making predictions?” from the Foundational Questions Institute and Fetzer Franklin Fund (a donor-advised fund of Silicon Valley Community Foundation). VN also acknowledges support from the Lee Kuan Yew Endowment Fund (Postdoctoral Fellowship).
\vspace{2ex}

\noindent\emph{Author contributions}\m PMR and RCH contributed equally to this work.
PMR, VN, and MG developed the theoretical formalism. RCH performed the numerical simulations and PMR performed the analytical calculations. All authors contributed to the writing of the manuscript. VN and MG supervised the project.

\bibliographystyle{quantum}
\bibliography{References}

\setcounter{secnumdepth}{3}
%\begin{widetext}
%\section{Supplementary Material}
\onecolumngrid
\appendix
\section{Decomposition of free energy in quantum patterns}
\label{app:FreeEnergyDecomp}

Consider a finite portion of the quantum pattern:
\be
\rho^{(1:L)} = \tr_{\bbZ \setminus \{ \ell \}_1^L }(\rho_{\dots A_{-1} A_0 A_1 \dots}) ~.
\ee
If each subsystem is non-interacting and the $\ell^\text{th}$ subsystem has 
%a local Hamiltonian with
a reference equilibrium Gibbs state $\gamma^{(\ell)}$,
then the nonequilibrium free energy for this portion of the pattern is given by
\begin{align}
\mathcal{F}^{(1:L)} 
&= 
F_\text{eq}^{(1:L)} 
+ \kB T \text{D} \Bigl[ \rho^{(1:L)} \| \bigotimes_{\ell = 1}^L \gamma^{(\ell)} \Bigr] \\
&= 
F_\text{eq}^{(1:L)}
+ \kB T \bigl[ \tr(\rho^{(1:L)} \ln \rho^{(1:L)}) - \sum_{\ell =1}^L \tr(\rho^{(\ell)} \ln \gamma^{(\ell)}) \bigr]
\label{eq:FreeEnergyDecomp2}
\\
&= 
 F_\text{eq}^{(1:L)} + 
 \kB T \underbrace{ \text{D} [ \rho^{(1:L)} \| \bigotimes_{\ell =1}^L \rho^{(\ell)} ]}_\text{total correlation}
+  \kB T \sum_{\ell =1}^L \text{D} [ \rho^{(\ell)} \|  \gamma^{(\ell)} ] 
~,
\end{align}
where $F_\text{eq}^{(1:L)}$
is the equilibrium free energy.

We recognize that 
$\text{D} [ \rho^{(1:L)} \| \bigotimes_{\ell =1}^L \rho^{(\ell)} ]$ is the \emph{total correlation}
within the quantum pattern,
while 
$\text{D} [ \rho^{(\ell)} \|  \gamma^{(\ell)}] $ is the 
\emph{local nonequilibrium addition to free energy}.
Each of these factors contributes uniquely to the free energy.  
When operating sequentially on 
each subsystem,
quantum pattern engines must leverage past information
to harvest the free energy 
in the correlations.

In the main text,
we suppose each non-interacting subsystem has the same local Hamiltonian, which implies that the reference Gibbs states are also all the same: $\gamma^{(\ell)} = \gamma$.
The general decomposition here shows that both quantum and classical correlations contribute to the extractable nonequilium addition to free energy.  However, in the main text, the quantum pattern is assumed to be classically-generated, despite having non-orthogonal states.
The inter-time quantum correlations are thus restricted to quantum discord, with no inter-time entanglement~\cite{Modi10Unified}.

\section{Memory-enhanced free energy}
\label{sec:MemoryEnhancementToFreeEnergy}

By self-consistency,
the local reduced state of $\rho_{\overleftrightarrow A}$ will simply be a probabilistic mixture of $\sigma^{(x)}$ in the form of $\braket{\xi_t}_{K_t} = \sum_{x \in \mathcal{X}} \gbm\pi  T^{(x)}  \mathbf{1} \, \sigma^{(x)} = \xi_0$.
The predicted quantum state $\xi_t$ thus has more free energy than the reduced state on average since
\begin{align}
\braket{\text{D}[\xi_t \| \gamma]}_{K_t} - \text{D}[\xi_0 \| \gamma]
&= 
\braket{\text{D}[\xi_t \| \gamma]}_{K_t} - \text{D}[\braket{\xi_t}_{K_t} \| \gamma ] 
\label{eq:RelEntConv}
\\
&= 
S(\braket{\xi_t}_{K_t}) - \braket{S(\xi_t)}_{K_t} 
\label{eq:EntropyConc}
\\
& \geq 0 
\label{eq:Nonneg}
~.
\end{align}
Thus, 
\begin{align}
\braket{\text{D}[\xi_t \| \gamma]}_{K_t} \geq \text{D}[\xi_0 \| \gamma] ~,
\end{align}
and so more work can be extracted on average when memory is leveraged to predict sequential quantum states.
The non-negativity of Eq.~\eqref{eq:Nonneg} can be seen either from the convexity of relative entropy in Eq.~\eqref{eq:RelEntConv}
or from the concavity of entropy in 
Eq.~\eqref{eq:EntropyConc}.

A nearly identical argument also
shows that the classically predicted state has more free energy than the average classical state:
$\braket{\text{D}[\xi_t^\text{dec} \| \gamma]}_{K_t} \geq \text{D}[\xi_0^\text{dec} \| \gamma]$, 
where the states of knowledge $K_t$ are now the classically induced ones.

\section{Synchronizing to a memoryful quantum source}\label{app:MSP}

Inferring the latent state of a known memoryful quantum source
allows maximal work extraction
when operating serially on the quantum states of the process.
The optimal state of knowledge, given a sequence of observations $\littleObs_1 \littleObs_2 \dots \littleObs_t$ obtained via interventions on the sequence of quantum systems $\sigma^{(x_1)}, \sigma^{(x_2)}, \dots \sigma^{(x_t)}$ is the conditional probability distribution induced by these interventions,
\begin{align}
\gbm\eta_t := \Pr(\St_t | \Obs_1 \dots \Obs_t = \littleObs_1 \dots \littleObs_t, \St_0 \sim \gbm\pi ) ~.
\label{eq:MxStDef}
\end{align}
The last condition $\St_0 \sim \gbm\pi$ means that the initial latent state of the generator is distributed as $\gbm\pi$.
This can be rewritten as 
$\gbm\eta_t = \sum_s \gbm\pi(s) \Pr(\St_t | \Obs_1 \dots \Obs_t = \littleObs_1 \dots \littleObs_t, \St_0 =s)$ for $t>0$.
Recall that $\gbm\pi = \gbm\pi \sum_{x \in \mathcal{X}} T^{(x)}$ is the stationary distribution over the states of the generator.
Thus, $\gbm\eta_0 = \gbm\pi$.

%First, w

If we 
introduce a new random variable $K_t$ to denote the optimally updated state of knowledge about the latent state of the
pattern generator, 
%of the classically-correlated quantum states
then we can replace the condition $\St_{t-1} \sim \gbm\eta_{t-1}$
with
$K_{t-1} = \gbm\eta_{t-1}$.
The condition on the state of knowledge is relevant to the extent that the choice of POVM is influenced by the state of knowledge.
We %assert 
remind the reader that in our framework
the POVM on the current
quantum output is chosen as a function of the state of knowledge $K_t$.

%Then the key insight in moving from Eq.~(20) to (21) is that: 
Note that the current
quantum output only depends on the current latent state of the process.  Accordingly,
the next observation---which is the outcome of the %[knowledge-informed] 
POVM on the current
quantum output---is conditionally independent of all previous outputs, given the current latent state and given the state of knowledge induced by all previous outputs.

We will now show that the optimal state of knowledge 
is recursive.
%can be updated iteratively.
I.e., we will show that: 
\begin{align}
\gbm\eta_t 
%& := \Pr(\St_t | \Obs_1 \dots \Obs_t = \littleObs_1 \dots \littleObs_t, \St_0 \sim \gbm\pi ) \\
& = 
\Pr(\St_t | \Obs_t = \littleObs_t, \St_{t-1} \sim \gbm\eta_{t-1} ) ~.
\label{eq:RecursiveMxSt}
\end{align}
%Where the equation follows the fact that quantum outputs are conditionally independent of all the other random variables given the latent state.
This follows from marginalizing over intervening latent states, employing Bayes' rule, and recognizing that the belief state $\gbm\eta_t$ is a function of the observations up to that time $\littleObs_{1} \dots \littleObs_t$. Starting from Eq.~\eqref{eq:MxStDef}, we find:
\begin{align}
\gbm\eta_t 
&:= 
\Pr(\St_t | \Obs_1 \dots \Obs_t = \littleObs_1 \dots \littleObs_t, \St_0 \sim \gbm\pi ) 
\nonumber \\
&= 
\sum_s 
\Pr(\St_t, \St_{t-1} = s | \Obs_1 \dots \Obs_t = \littleObs_1 \dots \littleObs_t, \St_0 \sim \gbm\pi ) 
\\
&= 
\frac{
\sum_s 
\Pr(\St_t, \Obs_t = \littleObs_t, \St_{t-1} = s | \Obs_1 \dots \Obs_{t-1} = \littleObs_1 \dots \littleObs_{t-1}, \St_0 \sim \gbm\pi ) }{\Pr(\Obs_t = \littleObs_t | \Obs_1 \dots \Obs_{t-1} = \littleObs_1 \dots \littleObs_{t-1}, \St_0 \sim \gbm\pi )}
\\
&= 
\frac{
\sum_s 
\Pr(\St_{t-1} = s | \Obs_1 \dots \Obs_{t-1} = \littleObs_1 \dots \littleObs_{t-1}, \St_0 \sim \gbm\pi )
\Pr(\St_t, \Obs_t = \littleObs_t | \Obs_1 \dots \Obs_{t-1} = \littleObs_1 \dots \littleObs_{t-1}, \St_0 \sim \gbm\pi , \St_{t-1} = s) }{\sum_{s'} \Pr(\Obs_t = \littleObs_t, \St_{t-1} = s' | \Obs_1 \dots \Obs_{t-1} = \littleObs_1 \dots \littleObs_{t-1}, \St_0 \sim \gbm\pi )}
\\
&= 
\frac{
\sum_s 
\gbm\eta_{t-1}(s)
\Pr(\St_t, \Obs_t = \littleObs_t | K_{t-1} = \gbm\eta_{t-1}, \St_{t-1} = s) }{
\sum_{s'} 
\gbm\eta_{t-1}(s')
\Pr(\Obs_t = \littleObs_t | K_{t-1} = \gbm\eta_{t-1}, \St_{t-1} = s') }
\\
&=
\Pr(\St_t | \Obs_t = \littleObs_t, K_{t-1} = \gbm\eta_{t-1} )
~.    
\end{align}
Hence, we have obtained Eq.~\eqref{eq:RecursiveMxSt}
from Eq.~\eqref{eq:MxStDef} as promised.

\begin{figure}[h]
\centering
\begin{tikzpicture}[node distance={15mm},main/.style = {draw, circle}] 
\node (1) {$\St_0$}; 
\node (2) [below right of =1]{$X_1$};
\node (3) [below of =2]{$\Obs_1$};
\node (4) [right of =1]{$\St_1$};
\node (5) [below right of =4]{$X_2$};
\node (6) [below of =5]{$\Obs_2$};
\node (7) [right of =4]{$\St_2$};
\node (8) [below right of =7]{$X_3$};
\node (9) [below of =8]{$\Obs_3$};
\node (10) [right of =7]{$\St_3$};
\node (11) [right of =8]{$\dots$};
\node (12) [below left of =3]{$K_0$};
\node (13) [right of =12]{$K_1$};
\node (14) [right of =13]{$K_2$};
\node (15) [right of =14]{$K_3$};

\draw[->](1) to (2);
\draw[->](1) to (4);
\draw[->](2) to (3);
\draw[->](2) to (4); 
\draw[->](4) to (5);
\draw[->](4) to (7);
\draw[->](5) to (6);
\draw[->](5) to (7);
\draw[->](7) to (8);
\draw[->](7) to (10);
\draw[->](8) to (9);
\draw[->](8) to (10); 
\draw[->](12) to (13);
\draw[->](13) to (14);
\draw[->](14) to (15);
\draw[->](12) to (3);
\draw[->](3) to (13);
\draw[->](13) to (6);
\draw[->](6) to (14);
\draw[->](14) to (9);
\draw[->](9) to (15);
\end{tikzpicture} 
 
\caption{Bayesian network showing the structure of conditional independencies among latent states $\St_t$ of the quantum source, the type $X_t$ of quantum state produced, the observable $\Obs_t$ attained from interaction, and the state of knowledge $K_t$ that influences the work extraction protocol.}
\label{fig:BayesNet}
\end{figure}
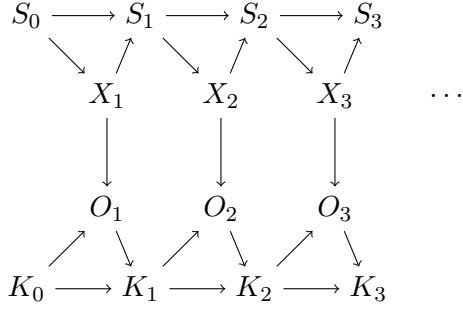

Further manipulations, using the rules of probability and the conditional independencies indicated in the Bayesian network depicted in Fig.~\ref{fig:BayesNet}, allow us to express the optimal state of knowledge in terms of both conditional work distributions and simple linear algebraic manipulations of the generative HMM representing the memoryful source. We find
\begin{align}
\gbm\eta_t 
& = 
\Pr(\St_t | \Obs_t = \littleObs_t, \St_{t-1} \sim \gbm\eta_{t-1} ) \\
& = 
\sum_{x \in \mathcal{X} }
\Pr(\St_t , X_t = x | \Obs_t = \littleObs_t, \St_{t-1} \sim \gbm\eta_{t-1} ) \\
& = 
\sum_{x \in \mathcal{X} }
\Pr( X_t = x | \Obs_t = \littleObs_t, \St_{t-1} \sim \gbm\eta_{t-1} )
\Pr(\St_t | X_t = x ,  \St_{t-1} \sim \gbm\eta_{t-1} ) \\
& = 
\sum_{x \in \mathcal{X} }
\Pr( X_t = x | \Obs_t = \littleObs_t, \St_{t-1} \sim \gbm\eta_{t-1} )
\frac{\gbm\eta_{t-1} T^{(x)}}{\gbm\eta_{t-1} T^{(x)}\mathbf{1}} \\
& = 
\frac{\sum_{x \in \mathcal{X} }
\Pr( X_t = x , \Obs_t = \littleObs_t | \St_{t-1} \sim \gbm\eta_{t-1} )
\gbm\eta_{t-1} T^{(x)} / \gbm\eta_{t-1} T^{(x)}\mathbf{1} }{
\sum_{x' \in \mathcal{X} }
\Pr( X_t = x' , \Obs_t = \littleObs_t | \St_{t-1} \sim \gbm\eta_{t-1} )
} \\
& = 
\frac{\sum_{x \in \mathcal{X} }
\Pr( \Obs_t = \littleObs_t | X_t = x, \St_{t-1} \sim \gbm\eta_{t-1} ) \, \gbm\eta_{t-1} T^{(x)} }{
\sum_{x' \in \mathcal{X} }
\Pr( \Obs_t = \littleObs_t | X_t = x', \St_{t-1} \sim \gbm\eta_{t-1} ) \, \gbm\eta_{t-1} T^{(x')} \mathbf{1}
}
~.
\end{align}

%If we 
%introduce a new random variable $K_t$ to denote the optimally updated state of knowledge about the latent state of the
%pattern generator, 
%then we can replace the condition $\St_{t-1} \sim \gbm\eta_{t-1}$
%with
%$K_{t-1} = \gbm\eta_{t-1}$.
%The condition on the state of knowledge is relevant to the extent that the choice of POVM is influenced by the state of knowledge.

\subsection{Using this to build a predictive work-extraction engine}

Rather than repeatedly calculating these ideal belief states on the fly for a specific realization of the process,
we can alternatively
systematically build up the set of all such belief states,
together with the
observation-induced transitions among them, to inform the design of an autonomous engine. 
There will be both a set of transient belief states and a set of recurrent belief states.
Both of these sets may be either finite or infinite.
In the case that only finitely many belief states are induced by observations,
we can explicitly build out the transition structure among them.  If there are infinitely many such states, then we would need to truncate unlikely states in the design of our finite physical engine~\cite{marzen17nearly}.

The physical memory system of our proposed engine should have at least one distinguishable state corresponding to every observation-induced belief state.
In fact, the memory must encode both the belief state and the most recent energy of the battery, so that conditioning on the new state of the battery is sufficient to supply the change in battery energy. 
These will likely be encoded with some finite precision, to avoid storing real numbers.
Conditioned on the state of the memory encoding $\gbm\eta$, the work extraction protocol will operate jointly on the quantum system, thermal reservoirs, and battery, to optimally extract work from the expected state $\xi = \sum_{x \in \mathcal{X}} \gbm\eta T^{(x)} \mathbf{1} \sigma^{(x)}$.

The subsequently observed work value $w$ 
uniquely updates
the memory from 
the state encoding $\gbm\eta$ to
the state encoding $\gbm\eta' = \frac{\sum_{x \in \mathcal{X} }
\Pr( W_t = w | X_t = x, \St_{t-1} \sim \gbm\eta ) \, \gbm\eta T^{(x)} }{
\sum_{x' \in \mathcal{X} }
\Pr( W_t = w | X_t = x', \St_{t-1} \sim \gbm\eta ) \, \gbm\eta T^{(x')} \mathbf{1}
}$.
Once the next quantum system arrives, the predictive quantum work extraction cycle begins again.

\section{Proof of Thm.~2: Work extraction, in the limit of zero entropy production}
\label{sec:ProofOfWorkStats}

Work extraction in the limit of zero entropy production
is important since it 
extracts all extractable work
%yields maximal possible work extraction 
from a quantum state.  It thus indicates the best possible scenario, against which other efforts can be compared.

In the limit of zero-entropy-production work extraction from $\rho^*$, the net unitary time evolution of the system--battery--baths supersystem must take a special form.
In particular, 
%as explained in App.~\ref{sec:ZeroEP},
the state of the battery will change deterministically when
the initial state of the system is an eigenstate $\ket{\lambda_n}$ of $\rho^* = \sum_n \lambda_n \ket{\lambda_n} \bra{\lambda_n}$,
almost-surely independent of the initial realization of the reservoirs.  
This implies that the net unitary time evolution
will be of the form
\begin{align}
U = \sum_{\varepsilon, n, r}
\Bigl( 
\ket{\varepsilon+ \wlambda[n]} \otimes 
 \ket{f_\varepsilon(n, r)} 
\Bigr)  \Bigl( 
\bra{\varepsilon} \otimes 
\bra{\lambda_n} \otimes 
\bra{r}
\Bigr)
\label{eq:UnitaryForm}
\end{align}	
for some $w^{(n)} \in \mathbb{R}$.
Above, $\ket{\varepsilon}$ and $\ket{\varepsilon + w^{(n)} }$
are energy eigenstates of the work reservoir, while $\ket{r}$ 
is an energy eigenstate of the thermal baths.
It will be useful in the following to note that
$\braket{f_\varepsilon(n, r) | f_\varepsilon(n', r')} = \delta_{n, n'} \delta_{r,r'}$
since unitary operations map orthogonal states to orthogonal states

The form of the unitary Eq.~\eqref{eq:UnitaryForm}
effectively assumes that the 
energy of the battery is well above its ground state.
Some interesting nuances 
have recently been explored
for batteries close to their ground state (see, e.g., Ref.~\cite{lipka21second}), 
which would affect the statistics of 
the work-extraction values,
but we
avoid that regime here 
to instead focus on the best-possible scenario.

%Note that a `work reservoir' is a subsystem that, when traced out, maintains unitary\m but not energy preserving\m time evolution on the rest of the universe.
%Accordingly, the joint state $\ket{f(\lambda, r)}$ 
%is a unitary %(although energy non-conserving) 
%transformation of $\ket{\lambda} \otimes \ket{r}$,
%which implies that $\braket{f(\lambda, r) | f(\lambda', r')} = \delta_{\lambda, \lambda'} \delta_{r,r'}$
%since unitary operations map orthogonal states to orthogonal states.

%(although its further details
%are irrelevant to the following).

One way to determine $\wlambda[n]$
is via the initial-state dependence of entropy production.
Let $\braket{\EP}_\rho$ denote the expectation value for entropy production, given initial system-state $\rho$, under the fixed work-extraction protocol optimized for $\rho^*$. In our case with a single heat bath at temperature $T$, 
the expected entropy production can be defined as usual as 
$\braket{\EP}_{\rho} = \bigl( \braket{\tilde{W}}_\rho - \Delta \mathcal{F}_t \bigr) / T$.
This is the entropy production for a fixed protocol operating on the initial state $\rho$,
where $\mathcal{F}_t$ is the nonequilibrium free energy at time $t$,
while $\Delta \mathcal{F}_t$ is the change in 
nonequilibrium free energy over the course of the protocol,
and $\tilde{W}$ is the work \emph{exerted}, which is just the negative of the extractable work~\cite{parrondo2015thermodynamics}.
%\PMR{add in the definition of entropy production}
Since all initial states map to $\gamma$ by the end of the work-extraction protocol,
we know
from Ref.~\cite{riechers2021initial} that
\begin{align}
	\braket{\EP}_{\sigma} - 	\braket{\EP}_{\rho^*} = \kB \text{D}[ \sigma \| \rho^*]  ~.
\end{align}	
In this case,
$\braket{\EP}_{\rho^*} = 0$
and 
$T \braket{\EP}_{\sigma} = \braket{\tilde{W}}_\sigma - \Delta \mathcal{F}_t = \braket{\tilde{W}}_\sigma + \kB T \text{D}[\sigma \| \gamma]$. 
%where $\mathcal{F}_t$ is the nonequilibrium free energy at time $t$,
%and $\tilde{W}$ is the work \emph{exerted}, which is just the negative of the extractable work.
Hence, with $\beta=(\kB T)^{-1}$,
\begin{align}
	\beta \braket{\tilde{W}}_\sigma 
	&=  \text{D}[ \sigma \| \rho^*] -  \text{D}[ \sigma \| \gamma ] \\
	&= \tr(\sigma \ln \gamma) - \tr(\sigma \ln \rho^*) ~.
\end{align}	
In particular, 
let $\sigma = \ket{\lambda_n} \bra{\lambda_n}$,
and note that 
$\ln \gamma = \ln (e^{-\beta H} / Z) = \beta(F-H)$.
This yields 
$\braket{\tilde{W}}_{\ket{\lambda_n} \bra{\lambda_n}} = F - \braket{\lambda_n| H | \lambda_n} - \kB T \ln \lambda_n$.
The deterministic work-extraction value, given initial pure state $\ket{\lambda_n}$,
must be the same as its expected value $\wlambda[n]  = -\braket{\tilde{W}}_{\ket{\lambda_n} \bra{\lambda_n}} $, and is thus given by
\begin{align}
	\wlambda[n] =  \braket{\lambda_n| H | \lambda_n} + \kB T \ln \lambda_n - F ~.
\end{align}

The probability of obtaining the work-extraction value $w$, given any input state 
$\sigma = \sum_{n, m} \ket{\lambda_n} \bra{\lambda_m} \braket{\lambda_n| \sigma | \lambda_m }$, 
can be calculated as
\begin{align}
\Pr(W = w | \sigma)
&=
\tr \Bigl[ \bigl( \ket{\varepsilon_0 + w} \bra{\varepsilon_0 + w} \otimes I \bigr) U \bigl( \ket{\varepsilon_0} \bra{\varepsilon_0} \otimes \sigma \otimes \ket{r} \bra{r}  \bigr) U^\dagger \Bigr] \\
&= \sum_{n, m}
  \braket{\lambda_n | \sigma | \lambda_m }
  \tr \Bigl[ \bigl( \ket{\varepsilon_0 + w} \bra{\varepsilon_0 + w} \otimes I \bigr) U \bigl( \ket{\varepsilon_0} \bra{\varepsilon_0} \otimes \ket{\lambda_n} \bra{ \lambda_m } \otimes \ket{r} \bra{r}  \bigr) U^\dagger \Bigr] \\
&= \sum_{n, m}
\braket{\lambda_n | \sigma | \lambda_m }
\tr \Bigl[ \bigl( \ket{\varepsilon_0 + w} \bra{\varepsilon_0 + w} \otimes I \bigr)  \bigl( \ket{\varepsilon_0 + \wlambda[n]} \otimes \ket{f_{\varepsilon_0}(n, r)} \bigr) \bigl( \bra{\varepsilon_0 + \wlambda[m]} \otimes \bra{f_{\varepsilon_0}(m, r)}  \bigr)  \Bigr]   \\
&= \sum_{n, m}
\braket{\lambda_n | \sigma | \lambda_m }
\delta_{w, \wlambda[n]}
\delta_{w, \wlambda[m]}
\braket{f_{\varepsilon_0}(m, r) | f_{\varepsilon_0}(n, r) }  \\
&= \sum_{n, m}
\braket{\lambda_n | \sigma | \lambda_m }
\delta_{w, \wlambda[n]}
\delta_{w, \wlambda[m]}
\delta_{n, m} \\
&= \sum_{n}
\braket{\lambda_n | \sigma | \lambda_n }
\delta_{w, \wlambda[n]}
\end{align}
independent of the initial energy state of the battery $\ket{\varepsilon_0}$, and almost-surely independent of the initial realization $\ket{r}$ of the thermal reservoirs in the probability theoretic sense. 
We see that $\Pr(W = w | \sigma) = 0$
unless $w \in \{ \wlambda[n] \}_n$.
The probability distribution over these
allowed work-extraction values is 
\begin{align}
\Pr(W = \wlambda[n] | \sigma)
&= \sum_{m}
\braket{\lambda_m | \sigma | \lambda_m }
\delta_{\wlambda[n], \wlambda[m]} ~.
\end{align}

When there is some entropy production,
the probability density of work extraction will have more diffuse peaks. However, for sufficiently low entropy production, the peaks will still be well separated and, so, effectively discrete for the purpose of Bayesian updating.

The above derivation is valid whether or not 
$\rho^*$ has degenerate eigenvalues.
Notably, the above sums are taken over the eigenstates and their associated eigenvalues, rather than summing over the eigenvalues directly.
%In particular, we can rewrite the spectral decomposition
%as $\rho^* = \sum_{n} \lambda_n \ket{n} \bra{n}$.
%With a slight change in notation, the allowed work-extraction values 
%can then be written as 
%\begin{align}
%	\wlambda[n] =  \braket{n| H | n} + \kB T \ln \lambda_n - F ~,
%\end{align}	
%which occur with probability
%\begin{align}
%\Pr(W = \wlambda[n] | \sigma)
%&= \sum_{m}
%\braket{m | \sigma | m }
%\delta_{\wlambda[n], \wlambda[m]} ~.
%\end{align}
%In the main text, we've avoided this notation
%so that eigenstates $\ket{n}$ and $\ket{m}$ are not confused with 
%computational-basis states.

\section{Power and inefficiency at rapid operation}

%As 
It is a familiar concept
in the design of any engine:
that
maximal thermodynamic efficiency requires sufficiently slow operation.
Clearly, this has implications for the power output of the engine~\cite{Curzon75Efficiency, Peterson19Experimental}.
However, the relaxation timescales of an engine depend on particular material properties of the system and baths, as well as the particular interaction Hamiltonian, so there is no implementation-independent timescale that determines the practical operation speed of an engine.

Nevertheless, we can apply rather general principles
to assess
%that we can assess to evaluate the scaling of 
%power with 
how power typically scales with 
increasingly fast operation.
For example,
%If each work-extraction step operates very quickly then, 
under assumptions of a Lindblad master equation,
%under the typical assumptions, 
there will be a contribution to entropy production (and a corresponding decrease in extracted work per operation) that scales as $1/\tau_0$, where $\tau_0$ is the duration of the work-extraction protocol~\cite{VanVu22Finite}. 
%[T.\ Van Vu and K.\ Saito,
%PRL 128 (1), 010602, 2022].
%The coefficient for this scaling is material dependent
Or, for unitary interactions with the bath,
a similar statement can be made but with
$\tau_0$ proportional to the number of interactions with bath degrees of freedom (i.e., the `circuit complexity')~\cite{Taranto23Landauer}. %independent of the time elapsed~\cite{Taranto23Landauer}.
In either case, we expect entropy production to scale as $\EP \approx c/(\kappa + \tau_0) \approx c/\tau_0$ for $\tau_0 \gg \kappa > 0$
where $c$ and $\kappa$ are implementation-dependent positive quantities.

In a fixed time $\tau$, the number of 
work-extraction operations $t$
determines the maximal allowed time 
$\tau/t \geq \tau_0$
for each work-extraction protocol.
Let $\braket{W_\text{ext}}$ be the steady-state work-extraction rate per operation in the limit of very slow operation (i.e., in the limit of infinitely many relaxation steps per work-extraction operation).
The power $P$ achieved by finite-time operation
is then sandwiched by
\begin{align}
\tfrac{t}{\tau} \braket{W_\text{ext}} 
\geq
P
=
\tfrac{t}{\tau} \bigl( \braket{W_\text{ext}} 
- c T / \tau_0
\bigr)
\geq
\tfrac{t}{\tau} \braket{W_\text{ext}} 
- c T / \tau_0^2
~.
\end{align}

%(or finite-interaction)

%If we focus on inefficiency of the engine due solely to this finite-time (or finite-interaction) effect, then we should expect the entropy production per work-extraction step
%to be 

We focus on the regime where each work-extraction protocol is of sufficiently long duration $\tau_0 \gg c/\kB$, such that $c T /\tau_0$ is negligible.
%, and most time is spent waiting for the next input.
In this regime, the power trivially scales with the number of operations per second, $P 
= \tfrac{t}{\tau} \braket{W_\text{ext}} 
\propto t/\tau$.
In the main text, 
we focus on results 
about $\braket{W_\text{ext}}$
obtained in
this simple regime, since it
highlights the thermodynamic role of correlations and quantum discord.

\section{The work extraction protocol of Skrzypczyk \textit{et al.}}
\label{supp:WEprot}

For self containment, here we give a brief summary of the Skrzypczyk \textit{et al.}\ work-extraction protocol~\cite{SSP14}, utilized in our numerical simulations.
Ref.~\cite{SSP14} should be consulted for further details.

The work extraction protocol of Ref.~\cite{SSP14},
%utilized in the numerical simulations, 
proceeds in two stages. 
The first stage (isentropically) puts the system into a mixture of energy eigenstates, using the battery to offset internal energy changes. In the second stage, the system is coupled with the ambient heat bath to slowly (in a sequence of $N$ steps) relax it into a thermal state, again using the battery to absorb all energy offsets.\\
Consider a system and a weight initially in an uncoupled state $\rho^* \otimes \rho_w$.
Let $\rho^*=\sum_n \lambda_n\ket{\lambda_n}\bra{\lambda_n}$ be a spectral decomposition of the system state $\rho^*$, with $\lambda_n \geq \lambda_{n+1}$. 
%(effectively ensuring that there is no population inversion in excited states). 
Step 1 of the work extraction protocol maps the eigenvectors $\ket{\lambda_n}$ into the energy eigenstates of the system, via the unitary operator
\begin{equation}
    V=\sum_n \ket{E_n}\bra{\lambda_n}\otimes \Gamma_{h_n},
\end{equation}
where $h_n=\bra{\lambda_n}H_s\ket{\lambda_n}-E_n$ accounts for the difference in energy in the two states and $\Gamma$ is an operator to raise the  potential energy of the weight. $V$ therefore conserves energy on average. The resultant state will be
\begin{equation}
    \rho_{\text{SW}}=\sum_n \lambda_n \ket{E_n}\bra{E_n}\otimes\Gamma_{h_n}\rho_w\Gamma_{h_n}^\dag
\end{equation}

In Step 2, the resultant state $\rho_{SW}$ will go through a sequence of transformation to reach the Gibbs state, $\gamma=\sum_n e^{-\beta E_n}/Z \ket{E_n}\bra{E_n}$.
%where $Z=\text{Tr}(e^{-\beta H})$ is the paritition function.
At each sub-step of the transformation, the relative probabilities of two levels will be adjusted 
by $\delta p$
towards the Gibbs distribution over energy eigenstates.
%towards the value of $e^{-\beta E_n}/Z$ by $\delta p$. 
Suppose we now focus on the occupation probability of the ground state and first excited state, $\ket{E_0}$ and $\ket{E_1}$. 
In the first 
sub-step of thermalization,
the system will interact with 
a bath qubit, $\rho_B=\frac{q_0}{q_0+q_1}\ket{0}\bra{0}+\frac{q_1}{q_0+q_1}\ket{1}\bra{1}$, where $q_0=\lambda_0-\delta p$ and $q_1=\lambda_1+\delta p$. A unitary transformation is then used to swap the occupation statistics of the bath qubit and the system qubit; in doing so, the battery's energy level rises or drops to conserve the total energy. The transformation can be described as
\begin{equation}
    \ket{E_0}_S\ket{1}_B\ket{x}_W \longleftrightarrow \ket{E_1}_S\ket{0}_B\ket{x+h}_W
\end{equation}
where $h=k_BT\log{\frac{q_0}{q_1}}-(E_1-E_0)$. The process then repeats itself by varying the value of $q_0$ and $q_1$ until $\rho_{SW}$ reaches a Gibbs state. It is provable that this protocol is able to extract all the free energy of the system up to $\mathcal{O}(\delta p^2)$.

\section{Simulation}
\label{supp:simulation}

\begin{figure}
    \centering
    \includegraphics[width=0.6\columnwidth]{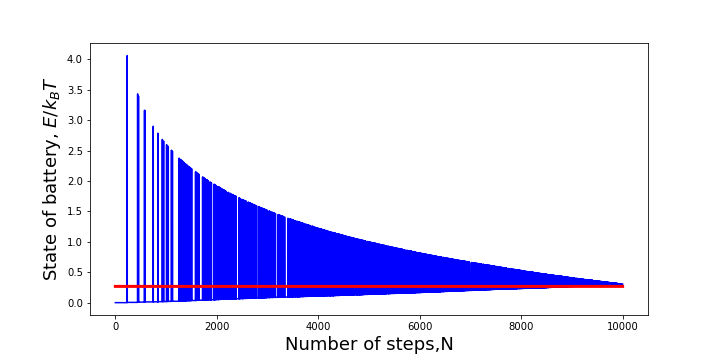}
    \caption{Quasistatic evolution of battery state over a finite number $N$ of bath interactions, under the work-extraction protocol from Skrzypczyk et al.\cite{SSP14} . Red line represents the total nonequilibrium addition to free energy present in the initial input state. }
    \label{fig:bat_evo}
\end{figure}

\begin{figure}
    \centering
    \includegraphics[width=0.6\columnwidth]{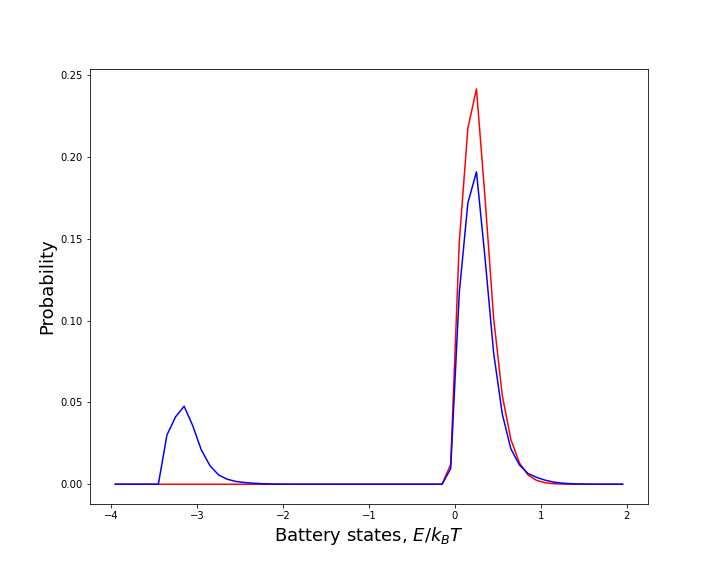}
    \caption{Probability distribution of work extracted, when using the the Skrzypczyk work-extraction protocol with a total of $N=22$ bath interactions. Red line represents the distribution when the protocol\m thermodynamically ideal for some mixed state very close to $\ket{0}$\m acts on the pure state $\ket{0}$. Blue represent the same protocol's work distribution when acting on a relatively non-orthogonal pure state $\ket{\psi}$, where fidelity between the two states is $F\bigl( \ket{0}, \ket{\psi} \bigr) = |\braket{0 | \psi}|^2 = 4/5$. }
    \label{fig:workdist}
\end{figure}

%\begin{figure}[t!] 
%\begin{center}
%         \includegraphics[width=0.5\columnwidth]{GM_Difference.png}
%         \caption{Contour plot of difference in average work extracted for Golden Mean process between memory assisted quantum protocol and memoryless quantum protocol. Red shows minimal improvement whereas blue region indicates high improvement.}
%     \label{fig:3D_GM}
%\end{center}
%\end{figure}
Here we elaborate on the the method of simulations. We considered a string of output with length $n=5000$ produced by the the perturbed-coin process. The possible emissions are $\sigma^{(0)}=\ket{0}\bra{0}$ and $\sigma^{(1)}=\ket{\psi}\bra{\psi}$ where $\ket{\psi}=\sqrt{r}\ket{0}+\sqrt{1-r}\ket{1}$. Here we considered two pure states\m this however is not necessary: mixed states can be used too. 

For the thermalization, the number of SWAP operations (between system and tailored baths) was chosen to be $N=200$ due to limitation in computational power. One can refer to Fig.~\ref{fig:bat_evo} to see that as $N\to\infty$ the protocol indeed extracts all the free energy from the system. 
After the extraction, the change in the battery system is measured and recorded. As mentioned in a previous section, if $N\to\infty$, the work distribution will converge to a set of $\delta$-functions. 
If $N$ is finite, the probability distribution of the work measured becomes more diffuse, as shown in  Fig.~\ref{fig:workdist}. 
The state of knowledge is updated via Bayesian inference, conditioned on the observed work value. 
%A new battery will be used in the subsequent iteration. 
This inference step is notably absent from the memoryless approach, where the state of knowledge effectively remains at the initial state of ignorance $\gbm \eta_t=\gbm\pi$.

For the classical protocol,
we assume that energetic coherences are 
inaccessible.
In this case, our simulations utilize 
the Skrzypczyk work-extraction protocol
tailored to the decohered state $\rho_t^* = \xi_t^\text{dec}$.
However, initial decoherence in any other basis,
besides $\xi_t$'s eigenbasis,
would likewise underperform compared to the memory-assisted quantum protocol. 

The overcommitment protocol differs from the rest as it tailors the extraction protocol for the state with the highest probability of emission. The probability of emission can be calculated from the state of knowledge $\gbm\eta_t T^{(x)}\gbm 1$. 
The graphs in Panels (c) and (d)
of 
Fig \ref{fig:Deg} display only positive work-extraction values on the vertical axis; hence most of the data points for overcommitment are not shown owing to its bad performance.

\section{Expected work extraction from the four approaches}
\label{app:expected work}

Here, we derive analytical expressions for the expectation value of work extraction from the various approaches compared in the main text.
We derive these expressions for the ($p$, $r$)-parametrized family of perturbed-coin processes 
of classically correlated quantum states discussed in the main text.

Recall that 
the quantum states $(\sigma^{(x)} )_{x\in \mathcal{X} }$ are the outputs of a Mealy HMM with labeled transition matrices $( T^{(x)} )_{x \in \mathcal{X}}$.  
An element of the labeled transition matrix $T_{s \to s'}^{(x)} = \Pr(X_t = x, \St_{t} = s' | \St_{t-1} = s)$ gives the joint probability of producing quantum state $\sigma^{(x)}$ and arriving at latent state $s'$, given that the HMM begins in state $s$.

For the perturbed-coin example, the HMM's labeled transition matrices are
\begin{equation}
    T^{(0)} = 
	\begin{bmatrix}
	1-p & 0 \\
	p & 0
	\end{bmatrix}
\qquad
\text{and }
\quad
	T^{(1)} = 
	\begin{bmatrix}
		0 & p \\
		0 & 1-p
	\end{bmatrix} ~.
\end{equation}
The stationary distribution over the latent states is $\gbm\pi=\left[\frac{1}{2},\frac{1}{2}\right]$ and the two different quantum states created are 
\begin{align}
    \sigma^{(0)} = \ket{0} \bra{0} =
 	\begin{bmatrix}
 		1 & 0 \\
 		0 & 0
 	\end{bmatrix}
 	\quad
 	\text{and }
 	\quad
 	\sigma^{(1)} = \ket{\psi} \bra{\psi} =
 	\begin{bmatrix}
 		r & \sqrt{r (1-r) }  \\
 		\sqrt{r (1-r)}  & 1 - r
 	\end{bmatrix} ~.
\end{align}
For any state of knowledge, $\gbm\eta_t=\left[ \frac{1}{2}+\epsilon_t, \frac{1}{2}-\epsilon_t \right]$ parameterized by $\epsilon_t \in [-\frac{1}{2},\frac{1}{2}]$, the induced expected state is 
\begin{align}
    \xi_t
	&
	= \rho^{(\epsilon_t)} :=
\sum_x  
	\begin{bmatrix}
		\tfrac{1}{2} + \epsilon_t &  \tfrac{1}{2} - \epsilon_t 
	\end{bmatrix}	
	T^{(x)} \mathbf{1} \sigma^{(x)} 
		= 
	\tfrac{1}{2}
	\begin{bmatrix}
		1 + r + \epsilon_t' \sqrt{1-r}  \,\, & \sqrt{r (1-r) }  - \epsilon_t' \sqrt{r}   \\
		\sqrt{r (1-r) }  - \epsilon_t' \sqrt{r}   \,\, & 1 - r - \epsilon_t' \sqrt{1-r}  
	\end{bmatrix} ~,
\end{align}
where $\epsilon' :=2 \epsilon(1-2p) \sqrt{1-r}\; \propto \epsilon$.

\subsection{Memory\hyp assisted quantum processing}\label{app:memfull}

In the memory-assisted quantum approach,
we utilize work-extraction protocols 
that are thermodynamically optimized for the expected quantum state
\begin{align}
    \rho_t^* = \xi_t = \rho^{(\epsilon_t)} ~.
\end{align}

Recall that the eigenvalues and eigenstates of $\rho_t^*$
play a prominent role in the work-extraction statistics.
We find that the eigenvalues of $\rho^{(\epsilon)}$ are
\begin{align}
    \lambda_{\pm}^{(\rho^{(\epsilon)})} 
	= \tfrac{1}{2} \pm \tfrac{1}{2}  \sqrt{r +  \epsilon'^2}~,
\end{align}
with corresponding eigenstates
\begin{align}
	\ket{\lambda_{\pm}^{(\rho^{(\epsilon)})}} = 
	2^{-1/2}
	\Bigl[ r + \epsilon'^2 \mp \bigl(r + \epsilon' \sqrt{1-r} \, \bigr) \sqrt{r + \epsilon'^2 }  \, \Bigr]^{-1/2}
	\begin{bmatrix}
		\sqrt{r(1-r)} - \epsilon' \sqrt{r}   \\
		- r -\epsilon' \sqrt{1-r}  \pm  \sqrt{r+\epsilon'^2}
	\end{bmatrix}  
	~.
\end{align}

For all times after $t=0$,
the update rule for belief states simplifies to the following
\begin{align}
    \gbm\eta_{t+1} \Bigr|_{ W_{t+1} = \wlambda[\pm] }
	& = 
	\frac{
		\sum_{x \in \mathcal{X} }
		 \bra{ \lambda_{\pm}^{(\xi_t)}  } \sigma^{(x)} \ket{ \lambda_{\pm}^{(\xi_t)}} \, 
		\gbm\eta_{t} T^{(x)} 
	}{
		\lambda_{\pm}^{(\xi_t)}
	} ~,
\end{align}
which can be expressed explicitly in terms of $p$, $r$, and $\epsilon_t$.

When $\epsilon_t=0$, we find that $\gbm\eta_{t+1}=\Bigl[ \frac{1}{2},\frac{1}{2} \Bigr] = \gbm\pi$. I.e., the stationary distribution is a fixed point for this dynamic over belief states.
Because of this, we break the initial symmetry by setting $\epsilon$ to a small non-zero value to obtain useful knowledge. 
In other words,
for the very first work-extraction protocol, we choose some $\rho_0^* \neq \xi_0$
to avoid an unstable fixed point of the update rule.
However, for all subsequent time steps, we choose $\rho_t^* = \xi_t$.

For the perturbed coin, the metadynamic of the belief state in the long run will yield two different results, depending on which regime the system is in, ``memory-apathetic regime" or ``memory-advantageous regime". 
%\begin{figure}
%    \centering
%    \includegraphics[width=0.7\columnwidth]{metadynamic.pdf}
%    \caption{Steady-state metadynamic among recurrent states of knowledge for the perturbed-coin process, in the memory-assisted quantum approach. Panel (a) represents the memory-apathetic
%    region of parameter space, where the sole recurrent state is the stationary distribution over latent states. 
%    Panel (b) represents the steady-state metadynamic throughout the memory-advantageous region of parameter space. The transition probabilities are governed by the eigenvalues of the induced expected state. }
%    \label{fig:metady}
%\end{figure}
The reason for this separation comes from the shape of their update function. For the memory-apathetic region, the update function has gradient less than unity, making $\epsilon=0$ an attractor. For the memory-advantageous region, the gradient of the update function exceeds unity, therefore making $\epsilon=0$ a repellor, at the same time two other points become part of a new attractor. 

%\begin{figure}
%    \includegraphics[width=\columnwidth]{figure_08.pdf}
%    \caption{Panel (a) shows the shape of the update functions in the memory-apathetic regime. Panel (b) shows the evolution of $\epsilon$ over time. Panel (c) shows the work-extraction values over 100 sequential time steps.}
%    \label{fig:agnostic}
%\end{figure}
%\begin{figure}
%    \centering
%    \includegraphics[width=\columnwidth]{figure_05.pdf}
%    \caption{Panel (a) shows the shape of the update functions in the memory-advantageous regime. Panel (b) shows the evolution of $\epsilon$ over time. Panel (c) shows the work-extraction values over 100 sequential time steps.}
%    \label{fig:advant}
%\end{figure}

In the long run,  transient belief states die out, leaving only the steady-state dynamics among the recurrent states of knowledge; any initial distribution over belief states generically converges to the stationary measure $\boldsymbol{\pi}_{\mathcal{K}}$. 
Hence the steady-state rate of work extraction is given by 
\begin{equation}
\lim_{t \to \infty}
    \langle{W_t}\rangle = \kB T \langle \text{D} [ \xi_t \| \gamma ]\rangle_{\Pr(K_t) = \boldsymbol{\pi}_{\mathcal{K}}} ~,
\end{equation}

The expected extracted work for the memory-apathetic region coincide with that of memoryless extraction and is given by
\begin{equation}
    \left\langle W^{\text{apathetic}}\right\rangle= \kB T\text{D}[\xi_0 \| \gamma]= \left\langle W^{\text{memoryless}} \right\rangle ~.
\end{equation}
On the other hand, in the regime where memory enhances the performance of the protocol, the stationary distribution over the two recurrent belief states $\gbm\eta$ and $\gbm\eta'$, with corresponding expected quantum states $\xi$ and $\xi'$, is
\begin{equation}
    \boldsymbol{\pi}_{\mathcal{K}} =\frac{1}{\lambda^{(\xi)}_++\lambda^{(\xi')}_+} [\lambda^{(\xi')}_+,\lambda^{(\xi)}_+] ~.
\end{equation}
%where $\lambda^{(\gbm\eta)}_+$ 
%is shorthand for $\lambda^{(g(\gbm\eta))}_+$
%with $g(\gbm\eta_t) := \gbm\eta_t 
%\begin{bmatrix} 1 \\ 0 \end{bmatrix}
%- 1/2 = \epsilon_t$.
Hence, the work extraction rate is given by 
\begin{equation}
    \left\langle W^{\text{advantage}} \right\rangle =\frac{\kB T }{\lambda^{(\xi)}_+ + \lambda^{(\xi')}_+} \Bigl( \lambda^{(\xi')}_+ \text{D}[\xi \| \gamma]+\lambda^{(\xi)}_+ \text{D}[\xi' \| \gamma] \Bigr) ~.
\end{equation}

\subsection{Classical approach}\label{app:class}
The derivation for the memory-assisted classical approach is similar to that of the memory-assisted quantum approach illustrated above. However rather than operating on the induced expected state $\xi_t$, the classical approach uses work-extraction protocols that are thermodynamically optimized for the decohered state 
\begin{equation}
    \rho_t^* = \xi_t^{\text{dec}}= 
	\frac{1}{2}\begin{bmatrix}
	1 + r + \epsilon_t' \sqrt{1-r}  \,\, & 0  \\
	0  \,\, & 1 - r - \epsilon_t' \sqrt{1-r}  
\end{bmatrix} ~.
\end{equation}

The eigenstates of $\rho_t^*$ are thus $\ket{0}$ and $\ket{1}$, independent of time in this case.
In the classical approach,
$\gbm\pi$ is no longer a fixed point of the belief-state update maps.
The transition probabilities between belief states are now given by
\begin{align}
\lambda_{\pm}^{(\xi_t^\text{dec})} = \tfrac{1}{2} \Bigl[ 1 \pm \bigl( r + \epsilon_t' \sqrt{1-r} \, \bigr) \Bigr] ~.
\end{align}

The metadynamic of belief in the classical case behaves as a reset processes. Unlike the quantum case with only two recurrent belief states, the 
classical protocol induces an infinite set of 
recurrent belief states.
%metadynamic of classical protocol is infinite in state space, w
To construct a finite-state autonomous engine, we could choose to truncate those states within some small $\delta$ distance from another recurrent state, or truncate belief states with negligible probability, with vanishing work-extraction penalty. 

We find that the work-extraction rate can again be computed by averaging the relative entropy\m now between the decohered expected state and thermal state\m over all recurrent states of knowledge:
\begin{align}
    \langle W_t^\text{classical} \rangle =
    \kB T \langle \text{D}[\xi_t^\text{dec} \| \gamma]  \rangle_{\Pr(K_t^\text{classical})} ~.
\end{align}

%\begin{figure}

%\begin{subfigure}{0.8\columnwidth}
%     \centering
%    \includegraphics{classicmeta.pdf}
%    \caption{}
%\end{subfigure}
%    \begin{subfigure}{0.8\columnwidth}
%         \includegraphics[width=\columnwidth]{classicmeta2.pdf}
%         \caption{}
%     \end{subfigure}
     
%\caption{Metadynamic of the belief state for the classical protocol. Panel (a) shows the exact transitions which resembles that of a reset process. Panel (b) shows the approximation used for computer simulation which allows us to get arbitrarily close to the full process in (a).}
%\label{fig:classicmeta}
%\end{figure}

\subsection{Overcommitment to the most likely outcome}\label{app:naive}

The ``overcommitment'' approach used for comparison in the main text bets exclusively on the most likely outcome in $\{ \sigma^{(x)} \}_x$.

The expected thermodynamic cost of misaligned expectations during work extraction can be quantified exactly via the relative entropy $\text{D} [ \rho_0 \| \alpha_0 ]$ between the actual input 
$\rho_0$ and the anticipated input $\alpha_0$ that the protocol is optimal for, if we assume that the final state is independent of the initial state~\cite{riechers2021initial, riechers2021impossibility}.  Hence, if we design the protocol for a pure state, but operate on a mixed state, we will encounter divergent thermodynamic penalties.

Accordingly, we can observe divergent thermodynamic costs when we design the Skrzypczyk work extraction protocol to be optimal for operation on a pure state.  

Using the Skrzypczyk protocol (with $N$ relaxation steps)
to extract work from the pure state bet upon,
we see that the first bath state swapped with the system for energy extraction is not exactly pure, but rather satisfies $\gamma_\text{B} = \Bigl( 1 - \frac{e^{-\beta E_0}}{N(e^{-\beta E_0} + e^{-\beta E_1})} \Bigr) \ket{0} \bra{0}
+ \frac{e^{-\beta E_1}}{N(e^{-\beta E_0} + e^{-\beta E_1})} \ket{1} \bra{1}$.
(Recall that $H$ is the Hamiltonian for the system, not of the bath.) Any purity of the actual input beyond this initial bath purity is wasted.
The input state leading to minimal entropy production under this protocol is thus %$\alpha_0 = 
a unitary rotation of $\gamma_\text{B}$.

Thus, for this use case of the Skrzypczyk protocol, the minimally dissipative state $\alpha_0$ becomes pure as $N \to \infty$.  As $N \to \infty$, we observe the battery's final expected energy diverging (but only logarithmically in $N$) to negative infinity, when this protocol acts on any other state.  I.e., 
$\langle W \rangle  \sim - \kB T \ln N$.
%$\langle W \rangle  \sim - \kB T (1-r) \ln N$.

More specifically, we can leverage 
Eqs.~\eqref{eq:WorkExtractionValues}
and \eqref{eq:WorkExtractionProbs}
to calculate the expected value of work
for the overcommitment approach.
%With
%$\lambda_\text{min} = \frac{e^{-\beta E_1}}{N(e^{-\beta E_0} + e^{-\beta E_1})}$,
%and 
%$\ket{\lambda_\text{min}} \perp \sigma^{(\text{argmax}_{x} \gbm\eta_t  T^{(x)}  \mathbf{1} )}$,
We find that 
\begin{align}
    \Pr(W=w^{(-)} | \sigma^{(\text{argmin}_{x} \gbm\eta_t  T^{(x)}  \mathbf{1} )}) = |\braket{1 | \psi}|^2 = 1-r ~.
\end{align}
With
$\lambda_- = \frac{e^{-\beta E_1}}{N(e^{-\beta E_0} + e^{-\beta E_1})}$,
$\wlambda[-] \sim - \kB T \ln N$, 
and 
$\text{min}_x  \gbm\eta_t  T^{(x)}  \mathbf{1}  \sim
\text{min}(p,1-p)$ when $\gbm\eta_t$ is close to either latent state,
we anticipate that the overcommited work penalty diverges as $- \kB T (1-r) \min(p, 1-p) \ln N$,
as observed.

Interestingly, for a finite number of bath interactions, some work can be extracted on average within certain regimes.  But other regions of parameter space would yield very negative work-extraction averages.

Unlike the other approaches,
the expectation value of work in the 
overcommitment approach cannot be written as a relative entropy.
Hence, whereas the other approaches were guaranteed to have non-negative work extraction on average,
the overcommitment approach enjoys no such guarantee of non-negativity.
Indeed in the limit of many bath interactions,
the overcommitment approach leads to infinitely negative work extraction.

\section{Quantum Information Processing Second Law}
\label{app:QIPSL}

Here, we derive the Quantum Information Processing Second Law (QIPSL),
which generalizes the classical IPSL 
of Refs.~\cite{mandal2012work,Boyd16Identifying, Boyd17Transient, Stopnitzky19Physical, Jurgens20Functional, Garner21Fundamental, He2022Information}
in several directions.

We consider the fundamental thermodynamic limits of a memoryful physical transducer that 
consumes one physical pattern and produces another.  Both the input and output are assumed to be stationary quantum stochastic processes.
Although much of the previous information-engines literature restricts itself to the energy-degenerate case (such that a `0' and `1' have the same energy), here we generalize to allow the Hamiltonian for each subsystem to be either degenerate or non-degenerate.
%We restrict ourselves to the case in which the Hamiltonian $H$ for each local element of the pattern is non-degenerate.

Denote the joint state of a $d_\text{\text{M}}$-dimensional memory and the length-$L$ pattern as $\rho_t^{(\text{M}, 1:L)}$, with corresponding reduced states of the memory $\rho_t^{(\text{M})}$
and of the pattern $\rho_t^{(1:L)}$.
If we denote $W_\tau^\text{ext} = \sum_{t=1}^\tau W_t$ as the net work extracted up to time $\tau$,
then the second law of thermodynamics 
%can be stated as an 
tells us that the change in nonequilibrium free energy 
$\mathcal{F}_t^{(\text{M}, 1:L)}$
upper bounds
the expectation value of extracted work when the environment is at the fixed temperature $T$~\cite{parrondo2015thermodynamics}.  We assume cyclic transformations where the 
joint Hamiltonian at time $\tau$ is the same as the the joint Hamiltonian at time $0$---although it may be modulated during the protocol to enable the work extraction.  
The equilibrium state $\boldsymbol{\gamma}$  for the joint pattern and memory
is thus the same at times $0$ and $\tau$.
For this reason, there is no
net change in equilibrium free energy.
Accordingly, the only change in nonequilibrium free energy is due to the change in the nonequilbrium addition to free energy---the quantum relative entropy between the actual state and the equilibrium state:
\begin{align}
\braket{W_\tau^\text{ext}} 
& \leq \mathcal{F}_0^{(\text{M}, 1:L)} - \mathcal{F}_\tau^{(\text{M}, 1:L)} \nonumber \\
& = \kB T \text{D}[ \rho_0^{(\text{M}, 1:L)} \| \boldsymbol{\gamma} ] - 
\kB T \text{D}[ \rho_\tau^{(\text{M}, 1:L)} \| \boldsymbol{\gamma} ] 
~.
\end{align}

We now make several assumptions relevant to energetically-efficient memoryful pattern manipulation:
\begin{enumerate}
\item
We will assume that the memory is fully energetically degenerate
so that unitary memory updates do not cost energy.
Then $\boldsymbol{\gamma} = (I/d_\text{M}) \otimes \gamma^{\otimes L}$ describes the equilibrium state of the joint supersystem.
\item
Since the memory is always initialized as a particular memory state $\rho_0^{(\text{M})}$, the memory and pattern are initially uncorrelated: $\rho_0^{(\text{M}, 1:L)} = \rho_0^{(\text{M})} \otimes \rho_0^{(1:L)}$.
\item
Through the sequence of deterministic updates, and equal (possibly zero) internal entropy of each memory state, the entropy of the memory is unchanging.
\end{enumerate}
Altogether, these features imply that the extractable work is upper bounded by
\begin{align}
\frac{\braket{W_\tau^\text{ext}}}{ \kB T} & \leq 
\text{D}[ \rho_0^{(1:L)} \| \gamma^{\otimes L}] - 
\text{D}[ \rho_\tau^{(1:L)} \| \gamma^{\otimes L}]
- \text{I}_\tau \\
%+ \text{I}_0 
&= - \text{I}_\tau
+ \Delta \text{S}(\rho_t^{(1:L)}) 
+ \Delta \sum_{\ell=1}^L \tr(\rho_t^{(\ell)} \ln \gamma) 
~,
\label{eq:AlmostQIPSL}
\end{align}
where 
$\text{I}_t
= \text{D}[ \rho_t^{(\text{M}, 1:L)} \| \rho_t^{(\text{M})} \otimes \rho_t^{(1:L)}  ]$
is the mutual information that has built up between the memory and the quantum pattern.
The operator $\Delta$ evaluates the difference of each function from time $t=0$ to time $t=\tau$,
such that
$\Delta f_t \equiv f_\tau - f_0$ for any function $f$ of $t$.
%In the main text, we furthermore assume that the memory is always initialized as $\rho_0^{(\text{M})}$, the memory and pattern are initially uncorrelated: $\rho_0^{(\text{M}, 1:L)} = \rho_0^{(\text{M})} \otimes \rho_0^{(1:L)}$, yielding $\text{I}_0 = 0$.

The classical IPSL, and the quantum generalization we derive,
address the scenario where
\emph{a physical machine scans unidirectionally along the pattern.}
%After $t$ time-steps, the first $t$ subsystems of the pattern have interacted with the machine, while the other subsystems remain unaltered:
%$\rho_t^{(t+1:L)} = \rho_0^{(t+1:L)}$.
After $\tau$ time-steps, the first $\tau$ subsystems of the pattern have interacted with the machine, while the other subsystems remain unaltered:
$\rho_\tau^{(\tau+1:L)} = \rho_0^{(\tau+1:L)}$.

More specifically, the IPSL assumes that both the input tape and output tape are stationary stochastic processes,
such that all marginal distributions of each (input or output) is shift invariant within the domain of the existing pattern.
We likewise consider the case that both input and output processes have a well-defined
von Neumann entropy rate $\entropyrate$ and $\entropyrate'$
respectively.
Each also has a \emph{myopic entropy rate} $\myopicentropyrate_\ell = \text{S}(\rho_0^{(1:\ell)}) - \text{S}(\rho_0^{(1:\ell-1)}) $
and 
$\myopicentropyrate_\ell' = \text{S}(\rho_\tau^{(1:\ell)}) - \text{S}(\rho_\tau^{(1:\ell-1)}) $ for $\ell \leq \tau$,
with $\myopicentropyrate_1 = \text{S}(\rho_0^{(1)})$
and $\myopicentropyrate_1' = \text{S}(\rho_\tau^{(1)})$.
The myopic entropy rate describes the 
convergence to 
irreducible randomness in pattern extension,
as progressively more subsystems are prefixed.
Relatedly, the \emph{myopic excess entropy} for the input process is 
$\mathcal{E}_\ell = \sum_{n=1}^\ell (\myopicentropyrate_n - \entropyrate)$,
with a similar relation for the primed output process.
Asymptotically, the excess entropy describes the quantum mutual information
between the past and future of a bi-infinite process:
$\lim_{\ell \to \infty} \EE_\ell = \EE = \lim_{\ell \to \infty} \text{I} (\rho_0^{(1:\ell)} ; \rho_0^{(\ell+1:2 \ell)})$.

%\emph{one stationary process is overwritten by another stationary process,
%as a physical machine scans along the pattern}.

With these definitions and assumptions in place, we can now 
expand Eq.~\eqref{eq:AlmostQIPSL} to
derive %and state 
the QIPSL:
\begin{align}
\frac{\braket{W_\tau^\text{ext}}}{ \kB T} 
& \leq 
- \text{I}_\tau
+ \Delta \text{S}(\rho_t^{(1:L)}) 
+ \Delta \sum_{\ell=1}^L \tr(\rho_t^{(\ell)} \ln \gamma) \nonumber \\
&=
- \text{I}_\tau
+ \Delta \text{S}(\rho_t^{(1:L)}) 
+ \Delta \sum_{\ell=1}^\tau \tr(\rho_t^{(\ell)} \ln \gamma) \\
&=
- \text{I}_\tau
+ \Delta \text{S}(\rho_t^{(1:L)}) 
+ \tau \Delta \tr(\rho_t^{(1)} \ln \gamma) 
~.
\label{eq:CloserToQIPSL}
\end{align}
Now, note that 
\begin{align}
\Delta \text{S}(\rho_t^{(1:L)}) 
&= 
\Delta \text{S}(\rho_t^{(1:\tau)})
+
\underbrace{\Delta \text{S}(\rho_t^{(\tau+1:L)})}_{=0}
- \Delta
\text{I}(\rho_t^{(1:\tau)} ; \rho_t^{(\tau+1:L)}) \\
&= - \text{I}(\rho_\tau^{(1:\tau)} ; \underbrace{\rho_\tau^{(\tau+1:L)}}_{=\rho_0^{(\tau+1:L)}}) + 
\underbrace{\Delta \text{S}(\rho_t^{(1:\tau)})}_{\tau ( \entropyrate' - 
\entropyrate ) +  \EE_\tau' - \EE_\tau}
+\underbrace{
\text{S}(\rho_0^{(1:\tau)}) + \text{S}(\rho_0^{(\tau+1:L)}) - \text{S}(\rho_0^{(1:L)})
}_{\EE_\tau + \EE_{L-\tau} - \EE_L} \\
&= 
- \text{I}(\rho_\tau^{(1:\tau)} ; \rho_0^{(\tau+1:L)})
+
\tau ( \entropyrate' - 
\entropyrate ) 
+  \EE_\tau'
- (\EE_L - \EE_{L-\tau})
~.
\end{align}

Thus, plugging this back in to Eq.~\eqref{eq:CloserToQIPSL}, 
we arrive at the 
\emph{Quantum Information Processing Second Law} 
(QIPSL): 
%\Rev{extensive derivation of eq 61 needed}
\begin{align}
\frac{\braket{W_\tau^\text{ext}}}{ \kB T} & \leq 
- \text{I}_\tau
- \text{I}(\rho_\tau^{(1:\tau)} ; \rho_0^{(\tau+1:L)})
+ \tau(\entropyrate' - 
\entropyrate)
+ \EE_\tau'
- (\EE_L - \EE_{L-\tau})
+ \tau
\Delta %\sum_{\ell=1}^\tau
\tr(\rho_t^{(1)} \ln \gamma) 
~.
\label{eq:QIPSL}
\end{align}
Under typical assumptions like finite memory, the time-extensive rate of work extraction during sequential quantum pattern manipulation is thus upper bounded by 
$\entropyrate' - 
\entropyrate + \Delta %\sum_{\ell=1}^\tau
\tr(\rho_t^{(1)} \ln \gamma) $, in units of $\kB T$:
\begin{align}
\lim_{\tau \to \infty} \frac{\braket{W_\tau^\text{ext}}}{ \tau \kB T} & \leq 
\entropyrate' - 
\entropyrate + \Delta 
\tr(\rho_t^{(1)} \ln \gamma) 
~.
\label{eq:QIPSL_rate}
\end{align}
Furthermore, if the Hamiltonian of each subsystem is fully degenerate, then $\Delta 
\tr(\rho_t^{(1)} \ln \gamma) = 0$, which would yield
$\lim_{\tau \to \infty} \frac{\braket{W_\tau^\text{ext}}}{ \tau \kB T} \leq 
\entropyrate' - 
\entropyrate $---analogous to the typical statement of the classical IPSL, but with von Neumann entropy density in place of the classical entropy density.

To be thermodynamically efficient,
it is known that
the classical machine needs to be  predictive or retrodictive in the case that it is consuming or generating a process, respectively~\cite{Boyd17Transient, garner2017thermodynamics, Boyd18Thermodynamics, Garner21Fundamental}.  From Eq.~\eqref{eq:QIPSL}, we now see that an analogous result remains true in the more general quantum case.

It is worth reflecting on the behavior of the contributions to Eq.~\eqref{eq:QIPSL}.
Notably, 
$\EE_L - \EE_{L-\tau} \geq 0$,
with 
$\EE_L - \EE_{L-\tau} \approx 0$ when $L-\tau \gg 0$,
and $\EE_L - \EE_{L-\tau} \to \EE_L \approx \EE$
as $\tau \to L$.
Meanwhile, the myopic excess entropy for the output process 
is a non-decreasing function of $\tau$
that approaches the past--future quantum mutual information of the output process
$\EE_\tau' \approx \EE'$ whenever $\tau \gg 0$.

\subsection{Implications for work extraction}

What are the implications for work extraction?
We note that after $t$ time-steps of work extraction from a quantum pattern, the first $t$ subsystems of the pattern have been brought to their equilibrium states, while the other subsystems remain unaltered:
$\rho_t^{(1:L)} = \gamma^{\otimes t} \otimes \rho_0^{(t+1:L)}$.
Hence 
$\text{I}(\rho_\tau^{(1:\tau)} ; \rho_0^{(\tau+1:L)}) = 0$,
and
$\text{I}_\tau = \text{D}[ \rho_\tau^{(\text{M}, \tau+1:L)} \| \rho_\tau^{(\text{M})} \otimes \rho_\tau^{(\tau+1:L)}  ]$
with
$\text{I}_\tau \to 0$
as $\tau \to L$.
Since the output process
produces an IID sequence of Gibbs states,
we have $\EE_t' = 0$,
while the entropy rate of the output equals the entropy of the Gibbs state
$\entropyrate' = \text{S}(\gamma)$.

Eq.~\eqref{eq:QIPSL} thus simplifies in this setting
to become
\begin{align}
\frac{\braket{W_\tau^\text{ext}}}{ \kB T} & \leq 
- \text{I}_\tau
- \tau 
\entropyrate 
- (\EE_L - \EE_{L-\tau})
-
\tau \tr(\rho_0^{(\ell)} \ln \gamma) \\
&=
- \text{I}_\tau
+ \tau [
\text{S}(\rho_0^{(\ell)}) - 
\entropyrate ]
- (\EE_L - \EE_{L-\tau})
+ \tau \text{D} [ \rho_0^{(\ell)} \| \gamma ] 
~,
\label{eq:QIPSLforWorkExtraction}
\end{align}
where the choice of $\ell \in \{ 1, 2, \dots , L\}$ is arbitrary since we assume a stationary quantum pattern.

For a long pattern,
the asymptotic work-extraction rate
is $\braket{W_t} = \lim_{\tau \to \infty} \frac{W_\tau^\text{ext}}{\tau}$.
When the work-extraction device has a finite memory, 
so that $\lim_{\tau \to \infty} \frac{\text{I}_\tau}{\tau} = 0$,
we thus find
that the steady-state work-extraction rate
is upper bounded by 
%the sum of the local nonequilibrium addition to free energy $\text{D}[\rho_0^{(\ell)} \| \gamma]$
%
%Dividing my $t$ yields an upper bound for the asymptotic rate
\begin{align}
\frac{\braket{W_t}}{\kB T} \, \leq \,
 \text{S} (\rho_0^{(\ell)}) - \entropyrate  +  \text{D} [\rho_0^{(\ell)} \| \gamma] \, \eqqcolon \, \frac{\wideal}{\kB T} ~.
\end{align}

%$\lim_{L \to \infty}
%\tfrac{1}{L} \braket{W_L^\text{ext}}$

When there is still plenty of input pattern remaining---i.e., when $L-\tau$ is sufficiently large such that $\EE_L - \EE_{L-\tau} \approx 0$---we find
$\braket{W_\tau^\text{ext}} \leq \tau  \wideal - \kB T \, \text{I}_\tau $.
%
%By the information-processing inequality, the investment cost $\text{I}_\tau$ is no more than the \emph{quantum excess entropy} $\mathcal{E}$,
%which is the quantum mutual information between past and future of a bi-infinite extension of the input process.
%However, the engine will only be able to harvest work at the ideal steady-state rate 
%if the memory stores all of the information from past inputs relevant for predicting future inputs, such that $I_t = \mathcal{E}$.
%Only then will the process be seen at the true entropy rate $\entropyrate$; otherwise it will appear more random than it really is, with the hidden structure evading extraction.
%
On the other hand, when 
the entire pattern has been consumed, 
$\text{I}_L = 0$ yet
$\EE_L - \EE_{0} = \EE_L \approx \EE $ for sufficiently large $L$,
leading to
\begin{align}
\braket{W_L^\text{ext}} \leq L \wideal - \kB T \mathcal{E} ~.
\end{align}
%The quantum excess entropy $\mathcal{E}$ is a thermodynamic investment for any ideal extractor---even one that operates non-locally.

\section{Non-Markovian generators}

Unlike the classical case,
the classical control symbols $X_t$ are hidden from direct observation when the process emits non-orthogonal quantum states.
The latent-state generators may thus be referred to as `doubly-hidden Markov models'.
Accordingly, even if the intermediary $X_t$ process is Markovian, this would
not directly imply a meaningful sense of 
quantum Markovianity of the outputs.
Nevertheless, there is some sense in which
processes with non-Markovian control outputs $X_t$ have more deeply hidden structure.

\begin{figure}
    \centering
    \includegraphics[width=0.65\columnwidth]{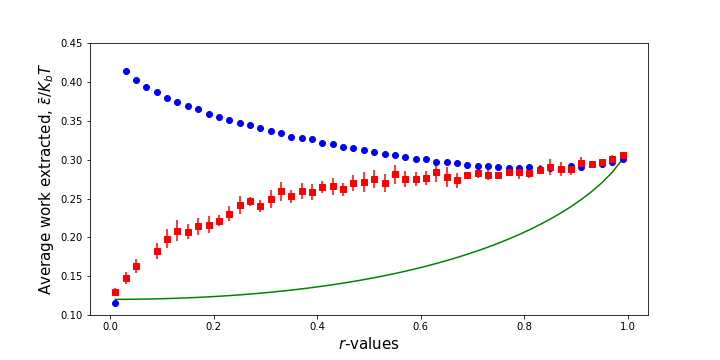}
         \caption{Comparison of average work extracted between 2-1 golden-mean and perturbed-coin processes, varying the nonorthogonality parameter $r$. Blue and red dots represent the memory-assisted quantum approach on golden mean and perturbed coin respectively; Green line represents the memoryless approach.}
         \label{fig:compare}
\end{figure}

To benchmark the performance of the memory-assisted protocol on a process with higher Markov order of the control symbols $X_t$, the 2-1 golden-mean process was chosen for comparison. The time-averaged density matrix of the memoryless approach for both models is kept the same, $\xi_0=\frac{1}{2}(\sigma^{(0)}+\sigma^{(1)})$. 
The comparison is shown in Fig.~\ref{fig:compare},
where we see that more work is extracted from the non-Markovian generator.
This example suggests that 
memory can become even more important for enabling work extraction from non-Markovian generators of quantum processes, since the extractable structure can be more deeply hidden.

%The reasoning is most transparent in the case of energy-degenerate Hamiltonians,
%where
%the maximal extractable work depends directly on 
%the von Neumann entropy rate 
%of each input process:
%\begin{align}
%\beta ( \mathcal{F}^{(1:L)} - F_\text{eq}^{(1:L)} ) / L
%&= \ln d
%- S_\text{vN} \bigl( \rho^{(1:L)} \bigr) / L
%\qquad \text{if } H \propto I ~,
%\end{align}
%as can be derived from Eq.~\eqref{eq:FreeEnergyDecomp2}.
%Here $S_\text{vN}(\rho) = -\tr(\rho \ln \rho)$ is the von Neumann entropy
%while $S_\text{vN} \bigl( \rho^{(1:L)} \bigr) / L$ is the von Neumann entropy rate.
%Unfortunately, there is no known closed-form expression for the von Neumann entropy rate from classical memoryfull generators of non-orthogonal quantum states.
%Nevertheless,
%non-Markovianity of the 
%control symbol $X_t$
%loosely suggests lower apparent von Neumann entropy rate, 
%and thus more extractable work,
%once longer lengths $L$ are taken into account.

%Unifilarity of the control symbol $x$ does not translate to any obvious closed-form solution for the von Neumann entropy rate.

%[Hence the difference in performance can be attributed primarily to the ability to retain memory. When we compare the result (Fig \ref{fig:compare}), the memory-enhanced work increases for processes that are non-Markovian (or higher Markov order).]
%{\color{red} PMR: The part in square brackets above is not rigorously justifiable.  It should be easy to come up with processes of any Markov order and mean quantum state with any work extraction curve.}

%\vspace{-1em}

\end{document}